\documentclass[a4paper,10pt]{article}
\usepackage[no-natbib-sort]{jheppub}

\usepackage[utf8]{inputenc}
\usepackage{etoolbox}
\usepackage{hhline}
\usepackage{array}
\usepackage{amssymb,amsmath,amsthm,amsfonts,graphicx}
\usepackage{latexsym}
\usepackage{bm}
\usepackage[]{hyperref}
\usepackage{cleveref}
\usepackage{nccmath}	%even more additional math symbols%
\usepackage{color}
\usepackage{ulem}
\usepackage{amssymb,amsbsy,enumitem}

\patchcmd{\maketitle}{\@fpheader}{}{}{}

\title{\boldmath Self-energy problem, vacuum polarization, and dual symmetry in Born-Infeld-type $U(1)$ gauge theories}

\author[a,b]{Ali Dehghani,}
\author[c]{Mohammad Reza Setare,}
\author[b,1]{and Soodeh Zarepour,\note{Corresponding author.}}

\affiliation[a]{Department of Physics and Biruni Observatory, College of Sciences, Shiraz University, Shiraz 71454, Iran}

\affiliation[b]{Department of Physics, University of Sistan and Baluchestan, Zahedan, Iran}

\affiliation[c]{Department of Physics, Campus of Bijar, University of Kurdistan, Bijar, Iran}

\emailAdd{ali.dehghani.phys@gmail.com}
\emailAdd{rezakord@ipm.ir}
\emailAdd{szarepour@phys.usb.ac.ir}

\abstract{We extensively explore three different aspects of Born-Infeld (BI) type nonlinear $U(1)$ gauge-invariant modifications of Maxwell's classical electrodynamics (also known as BI-type nonlinear electrodynamics) and bring some new perspectives on these theories. First, within the framework of exponential $U(1)$ gauge theory, it is explicitly proved that although the electric field at the location of the elementary point charges is not finite, but the total electrostatic field energy is finite. Motivated by this observation together with a wealth of evidence, we conjecture that all theories in 4-dimensional spacetime that belong to the BI family result in finite self-energy for elementary charged particles. In higher dimensions, it is found that the weak-field coupling limit of BI-type theories, which is identified as the weak field limit of effective Euler-Heisenberg (EH) theory, does not possess a regularizing ability to make the self-energy of a point charge finite, which implies that the conjecture may not hold for some BI-type theories. However, we explicitly prove that BI, logarithmic and exponential $U(1)$ gauge theories in arbitrary dimensions result in finite self-energy as well. Next, we classically study the problem of vacuum polarization effects and then systematically make a connection between all BI-type theories and QED. It is shown that all the BI-type theories classically predict the vacuum polarization effects, in which the final results are exactly in one-to-one correspondence with QED and the effective EH theory up to the leading order of corrections. Finally, we present a new, simple proof indicating that BI theory is the only theory with dual symmetry property (electric-magnetic duality) which reduces to the effective EH theory in the weak-field limit. The dual symmetry is broken when we demand an exact one-to-one correspondence in the weak-field limit with QED due to the vacuum polarization effects.}


\begin{document} 
\maketitle
\flushbottom

\section{Introduction}
\label{sec:intro}

The infinite self-energy of elementary point charges, e.g., charged leptons, has presented a major obstacle to classical physics. In order for resolving this problem, in 1934, Born and Infeld proposed a new classical field theory for $U(1)$ gauge field ``\textit{as a revival of the old idea of the electromagnetic origin of mass}" by postulating an upper bound for the purely electric field of elementary point charges like electron \cite{Born-Infeld1934}, whose Lagrangian is very similar with
that of Einstein's special relativity \cite{Einstein1905}. Today, this theory is called Born-Infeld (BI) nonlinear electrodynamics and it naturally eliminates the infinite classical self-energy of charged point particles, yielding a finite value for the total electrostatic field energy. The point-like nature of the elementary charged particles causes the infinity and the alternative way of resolving the self-energy problem is to consider just a nonzero minimal observable length\footnote{The significance of a minimal observable length in elementary particle physics was first understood by Heisenberg in ref. \cite{Heisenberg1938}. Such a minimal observable length could eliminate the ultraviolet divergences from the quantum field theory of the elementary particles \cite{MoayediSetare2012,Kempf1997,Hossenfelder2006}.} On the other hand, QED does not assist in resolving the infinite self-energy of elementary point charges \cite{Feynman1964}. For these reasons, modified theories of classical electrodynamics have received increased attention due to a number of various motivations, specially for resolving the infinite self-energy at the classical level. Remarkably, unlike the modified electrodynamics theories with a minimal length, the notion of point-like charge is still preserved in BI theory. In what follows, we shall restrict ourselves to BI family of $U(1)$ gauge theories, in which elementary charged particles are assumed to be point-like and size-less, in accordance with the primary assumption of the standard model of elementary particles.

The finiteness of self-energy of elementary point charges is probably one of the most important features of BI or any other nonlinear electrodynamics theory which is devised for this purpose. Although the BI theory historically is the first nonlinear theory of classical electrodynamics for this purpose, but it is quite possible that similar results can be obtained using the other types of Lagrangians. Another option is the logarithmic $U(1)$ gauge theory, proposed by Soleng in 1994 \cite{Soleng}, in which the corresponding Lagrangian has been constructed based on the logarithmic function and, consequently, the self-energy of point charges is always finite since there exists an upper limit on the electric field again  \cite{Gaete2014a}. Recently, the exponential form of nonlinear $U(1)$ gauge theory has also been presented in order to find a new class of asymptotic charged BTZ and Reissner–Nordström black hole solutions \cite{Hendi2012, Hendi2013}. However, in this theory, the electrostatic field at the position of the point charge still diverges but the divergence is much weaker than the Coulomb's inverse-square law, $\frac{1}{r^2}$. No investigation has yet been provided to address the self-energy issue in the exponential model of nonlinear $U(1 )$ gauge theory.

These theories (BI \cite{Born-Infeld1934}, logarithmic \cite{Soleng}, exponential \cite{Hendi2012}, etc \cite{Kruglov2016,DoubleLogaritmicNED2021}) share some common features especially in the weak-field coupling limit, leading to what might be called BI-type nonlinear theories of electrodynamics. Several BI-type Lagrangians are listed below
\begin{equation}
{\cal L}({\cal F}):\left\{ \begin{array}{l}
{{\cal L}_{{\rm{BI}}}}({\cal F}) = {\beta ^2}\left( {1 - \sqrt {1 + \frac{{2\cal F}}{{{\beta ^2}}}} } \right)\\
{{\cal L}_{\log }}({\cal F}) =  - {\beta ^2}\ln \left( {1 + \frac{{\cal F}}{{{\beta ^2}}}} \right)\\
{{\cal L}_{\exp }}({\cal F}) = {\beta ^2}\left( {{{\mathop{\rm e}\nolimits} ^{ - {\cal F}/{\beta ^2}}} - 1} \right)\\
\,\, \vdots 
\end{array} \right.
\end{equation}
where $\vdots$ denotes the other possible Lagrangians (see, e.g., refs. \cite{Kruglov2016,DoubleLogaritmicNED2021} and references therein), $\beta$ is the nonlinear parameter, and $\cal F$ is the usual Maxwell invariant as 
\begin{equation}
{\cal F} = \frac{1}{4}{F_{\mu \nu }}{F^{\mu \nu }} = \frac{1}{2}\left( {{{\bf{B}}^2} - {{\bf{E}}^2}} \right),
\end{equation}
in which ${F_{\mu \nu }}$ is the Faraday tensor. These models have the following expansion in the weak-field limit (large enough $\cal \beta$)
\begin{equation} \label{asymp-BI}
{\left. {{\cal L}({\cal F})} \right|_{{\rm{large}}\,\beta }} =  - {\cal F} + {a_1}\frac{{{{\cal F}^2}}}{{{\beta ^2}}} - {a_2}\frac{{{{\cal F}^3}}}{{{\beta ^4}}} + {a_3}\frac{{{{\cal F}^4}}}{{{\beta ^6}}} + O\left( {{\beta ^{ - 8}}} \right),
\end{equation}
where $a_i$’s are some positive constants and in our notation we always have $a_1=\frac{1}{2}$. Despite a series of similarities between these theories, there are some differences between them which depend on the functional form of the Lagrangian density. For example, the vacuum birefringence phenomenon is present in logarithmic and exponential models of nonlinear electrodynamics, while it is absent in BI theory \cite{Gaete2014a,Gaete2014b}. This is of importance since QED predicts the vacuum birefringence as well \cite{VB2013a,VB2013b,VB2017} and empirical evidence for this phenomenon has now been found \cite{VB-exp-2013,VB-exp-2014,VB-exp-2016}. On the other hand, the dual symmetry (the electric-magnetic duality) seems to be only present in BI theory \cite{Gibbons1995,Gibbons1996} but not in the other models of BI-type $U(1)$ gauge theories.\footnote{In section \ref{sec:duality}, we have presented a new, simple proof for this claim. This proof is based on the fact that BI-type theories generally have a weak-field limit the same as eq. (\ref{asymp-BI}) or more generally the same as eq. (\ref{asymp-BI-EH}). Otherwise, leaving aside this physical assumption, another nonlinear extensions of Maxwell's Lagrangian density with dual symmetry property may still be found \cite{ModMax2020PRD,ModMax2021JHEP,Babaei2016,Babaei2021}. For more details on this issue, see sect. \ref{sec:duality}.}

The classical theories of nonlinear electrodynamics cannot explain why the nonlinearity occurs in electromagnetic fields, instead it is assumed that electromagnetic fields are inherently nonlinear. However, there exists some evidence in support of a nonlinear effective theory for electromagnetic phenomena, so that the Maxwell's classical electrodynamics might emerge in the weak-field coupling limit. In fact, Heisenberg and Euler have shown that the Maxwell equations even in the vacuum have to be exchanged by more fundamental $U(1)$ gauge theories of nonlinear electrodynamics in order to explain the vacuum polarization effects \cite{HeisenbergEuler1936}. This theory takes into account vacuum polarization effects to one loop (the so-called photon self-energy diagrams) and effectively describes the nonlinear dynamics of electromagnetic fields. In the weak-field coupling limit, the EH Lagrangian reduces to
\begin{equation} \label{EH Lagrangian}
{{\cal L}_{{\rm{EH}}}}({\cal F},{\cal G}) =  - {\cal F} + \frac{{2{\alpha ^2}}}{{45m_e^4}}\left( {4{{\cal F}^2} + 7{{\cal G}^2}} \right) - \frac{{64\pi {\alpha ^3}}}{{315m_e^8}}\left( {16{{\cal F}^3} + 26{\cal F}{{\cal G}^2}} \right) + ...
\end{equation}
where $\alpha$ is the fine-structure constant and $m_e$ is the electron mass. Henceforth, the weak-field limit of EH theory is referred to as EH theory by us, which is common in the literature. In the above Lagrangian, the Maxwell invariant $\cal G$ is defined as
\begin{equation} \label{invariant G}
{\cal G} = \frac{1}{4}{F_{\mu \nu }}{}^*{F^{\mu \nu }} = {\bf{E}}.{\bf{B}},
\end{equation}
where ${}^*{F^{\mu \nu }} (= \frac{1}{2}{\varepsilon ^{\mu \nu \alpha \beta }}{F_{\alpha \beta }})$ is the Faraday tensor's Hodge dual. On the other hand, the quantum mechanical nonlinearity of electromagnetic fields, which leads to scattering of light-by-light (also known as Halpern scattering, first predicted in refs. \cite{Halpern1934,Euler-Kockel1935,Euler1936}), can effectively be described by BI-type nonlinear electrodynamics as well. In fact, the classical electrodynamics is in agreement with QED at the tree level \cite{KlauberQFT,SchwartzQFT} and since there is no tree-level contribution to the process of photon-photon scattering in QED, so it is understandable why classical electrodynamics cannot effectively explain light-by-light scattering. Today, it is a well-known fact that light-by-light scattering ($\gamma \gamma \to \gamma \gamma$), which is a very rare phenomenon, can be classically described within the context of nonlinear electrodynamics theories \cite{Thomas1936,Schrodinger1941,Davila2014,Ellis2017,Rebhan2017} and quantum mechanically by radiative corrections in QED \cite{KlauberQFT, SchwartzQFT,MandlShaw}. The strong evidence for this process has been recently reported after a long search by the ATLAS Collaboration \cite{ATLAS2017}.

As seen, BI-type models were introduced with various motivations but another important motivation for considering them comes from the string theory \cite{StringBI1,StringBI2,Gross1987,StringBI3,StringBI4}. In fact, the low-energy limit of $E_8 \times E_8$ heterotic string theory reduces to an effective field theory including a quadratic Maxwell invariant term as well as the Maxwell Lagrangian. (In principle, any BI-type theory involves all higher-derivative corrections of the $U(1)$ gauge field.) On the other hand, the dynamics of electromagnetic fields on the world-volumes of D-branes are governed by BI theory. Since both the Einstein-Hilbert action (in an anti-de Sitter background) and BI effective action appear at low-energy limit of string theory models, coupling nonlinear electrodynamics to gravity has been found various applications in the literature, especially in the contexts of effective field theories, gauge/gravity duality and black hole physics \cite{Soleng,Hendi2012,Hendi2013,Gibbons1996,StringBI1,StringBI2,StringBI3,StringBI4,Beato1999,Yajima2001,Zumino2008,Munoz2009,Miskovic2011a,Miskovic2011b,Zhao2013,Ruffini2013,Hendi2015Dehghani,Dehyadegari2016,Gulin2017,Falciano2019,Cremonini2019,Zarepour2021,Amirabi2021}.

For these reasons, we regard BI-type theories as a subclass of nonlinear $U(1)$ gauge theories that merit further exploration. We intend to extensively explore three different aspects of these theories, including the self-energy problem, the vacuum polarization effects and the dual symmetry. Usually, the aforementioned issues are investigated case by case for each BI-type model. This seems inevitable when the strong-field limit becomes significant (e.g., in the case of the self-energy problem). Here, we address the three mentioned issues in a general way in order to gain more insight into BI-type $U(1)$ gauge theories, yielding a better understanding of the outcomes of these theories. Furthermore, the four cases of BI, EH, logarithmic and exponential models will explicitly be discussed as the most important examples of nonlinear models of classical electrodynamics.\footnote{So far the investigations for nonlinear models of classical electrodynamics have concentrated more on the cases of BI, EH and logarithmic theories, but other BI-type models including exponential nonlinear model have been considerably less explored.}

Taking the above mentioned motivations into account, this paper is organized as follows. In section \ref{sec:BI-theories}, by considering BI-type theories involving the Maxwell invariant $\cal G$, we will develop the basic tools which are necessary for studying these theories in the next sections. The implications of the analysis are that one does not need to investigate these problems case by case for each BI-type nonlinear electrodynamics theory. In section \ref{sec:self-energy}, we first address the self-energy problem of point charges in the unstudied case of exponential electrodynamics and then extend our consideration to higher dimensions for all the BI-type theories under consideration. Actually, this will be the first study of the self-energy problem in higher dimensions within the context of nonlinear $U(1)$ gauge theories and a series of general conclusions will be presented for this family of theories. Afterwards, in section \ref{sec:vacuum polarization}, we show how all the BI-type $U(1)$ gauge theories can systematically explain the vacuum polarization effects exactly the same as QED. In section \ref{sec:duality}, we reconsider the electric-magnetic duality invariance in the BI family of nonlinear $U(1)$ gauge theories and bring a new perspective on this issue. Finally, in section \ref{sec:conclusion}, we summarize the main results.

\section{BI-type nonlinear $U(1)$ gauge theories} \label{sec:BI-theories}

In this section, we will develop the basic framework that we need to study the BI family of $U(1)$ gauge theories in the next sections. A useful notation will be introduced in sections \ref{sec:2-1} and \ref{sec:field eqs} in order to study BI-type theories within a general method. We use the Gaussian (cgs) units and also set $c=\hbar=1$, but these constants will be restored in some sections for numerical purposes. All physical constants used in this paper are taken from PDG \cite{PDG2020}.

\subsection{Building a nonlinear Lagrangian} \label{sec:2-1}

Nonlinear electrodynamics in the weak-field limit (or equivalently for negligible nonlinearity) and up to the first-order approximation must agree with Maxwell classical electrodynamics. This is a very important constraint that makes the new theory viable and the physical outcomes of Maxwell's theory are guaranteed to this extent. On the other hand, we are interested in a more generalized BI-type family of $U(1)$ gauge theories by adding the Maxwell invariant $\cal G$, eq. (\ref{invariant G}), to the action. When ${\bf {B}} = 0$, the second Maxwell invariant ($\cal G$) vanishes trivially, so it has no role in self-energy problem. In general, a nonlinear electrodynamics Lagrangian density in vacuum is built up from the electromagnetic invariants, $\cal F$ and $\cal G$, i.e., ${\cal L} = {\cal L}\left( {{\cal F},{\cal G}} \right)$. In this work, we restrict ourselves to the BI-type nonlinear electrodynamics \cite{Born-Infeld1934,Soleng,Gaete2014a,Hendi2012,Hendi2013,Kruglov2016,DoubleLogaritmicNED2021,Gaete2014b,Davila2014,Ellis2017,Rebhan2017,Miskovic2011a,Miskovic2011b,Zhao2013,Hendi2015Dehghani,Dehyadegari2016,Gulin2017,Zarepour2021,Kruglov2017,Akmansoy2018} for which the Lagrangian density is an analytic function of $\cal F$ and $\cal G$, therefore, it can be written as a power series of the electromagnetic invariants as
\begin{equation}
{\cal L} = \sum\limits_{p,q} {{a_{pq}}{{\cal F}^p}{{\cal G}^q}}  = {a_{10}}{\cal F} + {a_{01}}{\cal G} + {a_{11}}{\cal F}{\cal G} + {a_{20}}{{\cal F}^2} + {a_{02}}{{\cal G}^2} + ... , \quad ({a_{10}} =  - 1)
\end{equation}
where the second term is neglected due to Bianchi identity.\footnote{Of course, there are various types of nonlinear electrodynamics which do not satisfy the analyticity condition and do not reduce to the weak-field limit of BI and EH theories. As an example, we can speak of the conformal nonlinear electrodynamics with applications in black holes and AdS/CFT contexts for which the conformal invariance is maintained \cite{NED-conformalI1,NED-conformalI2,NED-conformalI3,NED-conformalI4,NED-conformalI5}. These models are beyond the scope of the present paper.} There are two physical reasons for considering the Maxwell invariant $\cal G$ in the action:
\begin{itemize}
	\item The Maxwell invariants $\cal F$ and $\cal G$ are two fundamental invariants that can be used to construct all possible invariants in electrodynamics (in four-dimensions). Therefore, any generalization of Maxwell's Lagrangian reduces the possibilities to powers or functions of the Maxwell invariants $\cal F$ and $\cal G$ \cite{LandauLifshitzBook}.
	
	\item It is well understood that the Maxwell invariant $\cal G$ is essential for describing the vacuum polarization effects in an effective field theory \cite{HeisenbergEuler1936}. In fact, in section \ref{sec:vacuum polarization}, it will be shown that the inclusion of the Maxwell invariant ${\cal G}$ in the theory can induce anisotropic polarization of the classical vacuum not only for EH theory but also for all BI-type theories, which is exactly similar to the result of QED for vacuum polarization effects up to the leading order of corrections (see more details in section \ref{sec:vacuum polarization}).
\end{itemize}
For these reasons, following Born and Infeld \cite{Born-Infeld1934}, we shall bring the Maxwell invariant $\cal G$ into our considerations by use of the following replacement
 \begin{equation} \label{S invariant}
{\cal F}\,\, \to \,\,{\cal S} = {\cal F} - \frac{{\gamma {{\cal G}^2}}}{{{\beta ^2}}}.
 \end{equation}
This extension is slightly more general than that of Born and Infeld (i.e., $\gamma = \frac{1}{2}$ \cite{Born-Infeld1934}) since the (dimensionless) coefficient $\gamma$ can also vary here. (This generalization has been already used in order for studying some specific models, e.g., see refs. \cite{DoubleLogaritmicNED2021,Kruglov2017}.) The primitive BI Lagrangian was a function of ${\cal S} = {\cal F} - \frac{{\cal G}^2}{2\beta^2}$ and this choice stems from the principle of least action and the assumption that the action should be scalar invariant. (The procedure is well explained in \cite{Born-Infeld1934}.) This choice also ensures that the BI-type models and their Taylor expansions will be invariant under parity transformation, i.e., ${\cal L}({\cal F},{\cal G})={\cal L}({\cal F},-{\cal G})$, and all of them can be expanded in powers of ${\cal G}^2$ in certain limits (for example, in the weak- or strong-field limits). Therefore, one can naturally generalize other BI-type nonlinear models using the replacement (\ref{S invariant}) and some explicit examples of such Lagrangian densities may be written as
\begin{equation} \label{BI-type theories}
{\cal L}({\cal F},{\cal G}):\left\{ \begin{array}{l}
{{\cal L}_{{\rm{BI}}}}({\cal F},{\cal G}) = {\beta ^2}\left( {1 - \sqrt {1 + \frac{{2\cal S}}{{{\beta ^2}}}} } \right)\\
{{\cal L}_{\log }}({\cal F},{\cal G}) =  - {\beta ^2}\ln \left( {1 + \frac{{\cal S}}{{{\beta ^2}}}} \right)\\
{{\cal L}_{\exp }}({\cal F},{\cal G}) = {\beta ^2}\left( {{{\mathop{\rm e}\nolimits} ^{ - {\cal S}/{\beta ^2}}} - 1} \right)\\
\, \vdots 
\end{array} \right.
\end{equation}
where $\vdots$ denotes the other possibilities.\footnote{It should be clear from the Lagrangians of BI and logarithmic models that there is an upper limit on the electric field, i.e., $E_{\rm{BI}}=\beta$ and $E_{\rm{log}}={\sqrt {2}} \beta$.} Using this simple replacement, generally, the weak-field expansion of BI-type theories involving the Maxwell invariant $\cal G$ take the following form
\begin{eqnarray} \label{asymp-BI-EH}
{\cal L}({\cal F},{\cal G}) =  &-& {\cal F} + \frac{1}{{{\beta ^2}}}\left( {{a_1}{{\cal F}^2} + \gamma {{\cal G}^2}} \right) - \frac{1}{{{\beta ^4}}}\left( {{a_2}{{\cal F}^3} + 2{a_1}\gamma {\cal F}{{\cal G}^2}} \right) \nonumber \\
&+& \frac{1}{{{\beta ^6}}}\left( {{a_3}{{\cal F}^4} + 3{a_2}\gamma {{\cal F}^2}{{\cal G}^2} + {a_1}{\gamma ^2}{{\cal G}^4}} \right) + O\left( {{\beta ^{ - 8}}} \right),
\end{eqnarray}
where $a_1 = \frac{1}{2}$ and the rest of $a_i$'s depend on the functional form of the nonlinear Lagrangian. By examining the Taylor expansion for other BI-type models, it is concluded that all the BI-type Lagrangians have the same mathematical form in the weak-field limit as eq. (\ref{asymp-BI-EH}), which is the natural result of the Taylor expansion of basic functions. This is a well-known, interesting result in comparison with the weak-field limit of EH Lagrangian (\ref{EH Lagrangian}) since both of them have the same mathematical form, meaning that, in all cases, the same electromagnetic invariants appear. So, we conclude that all BI-type models of nonlinear electrodynamics in the weak-field limit reduce to the weak-field limit of EH theory (\ref{EH Lagrangian}).\footnote{A possible alternative Lagrangian density based on the exponential function has been proposed in ref. \cite{Kruglov2016}. In accordance with our notation in this paper, the Lagrangian density of this model may be written as
\begin{equation}
{{\cal L}_{\exp }} =  - {\cal F}{{\rm{e}}^{-\frac{\cal F}{2\beta ^2}}},
\end{equation}
where, with the replacement (\ref{S invariant}), its weak-field expansion exactly matches with eq. (\ref{asymp-BI-EH}) and also with that of ref. \cite{Hendi2012}. Although all the general results of this paper are valid for this model as well, it seems to be difficult to work with it. In ref. \cite{DoubleLogaritmicNED2021}, the authors have also introduced a new nonlinear electrodynamics model given by (again, in accordance with our notation in this paper)
\begin{equation}
		{{\cal L}_{{\rm{Dlog}}}} = \frac{{{\beta ^2}}}{3}\left[ {\left( {1 - \sqrt { - \frac{{3{\cal S}}}{{{\beta ^2}}}} } \right)\ln \left( {1 - \sqrt { - \frac{{3{\cal S}}}{{{\beta ^2}}}} } \right) + \left( {1 + \sqrt { - \frac{{3{\cal S}}}{{{\beta ^2}}}} } \right)\ln \left( {1 + \sqrt { - \frac{{3{\cal S}}}{{{\beta ^2}}}} } \right)} \right],
\end{equation}
which reduces to eq. (\ref{asymp-BI-EH}) in the weak-field limit (again, with $a_1 = \frac{1}{2}$). Note that here the nonlinear parameter $\beta$ is different from the one in ref. \cite{DoubleLogaritmicNED2021}. Using the substitution $\beta^2 \to \frac{3}{2 \beta}$, the above Lagrangian is converted to its original form proposed in ref. \cite{DoubleLogaritmicNED2021}.  In addition, the first logarithmic part of this Lagrangian, i.e., ${\ln \left( {1 - \sqrt { - \frac{{3{\cal S}}}{{{\beta ^2}}}} } \right)}$, implies a maximum value for the electric field as $E_{\rm{Dlog}}=\sqrt{\frac{2 }{3}}\beta$.} Hereafter, the weak-field limit of EH theory up to the leading order of corrections is simply referred to as EH theory, which determines the weak-field behavior of BI-type models. In order to compare the physical results of EH theory with the results of other theories of the BI-type family, the EH Lagrangian density (\ref{EH Lagrangian}) may be written as
\begin{equation} \label{EH}
{{\cal L}_{{\rm{EH}}}}({\cal F},{\cal G}) =  - {\cal F} + b\left( {{{\cal F}^2} + \gamma {{\cal G}^2}} \right) + O\left( {{b^2}} \right),
\end{equation}
where the EH parameter $b$ is related to the BI nonlinear parameter $\beta$ as $b=\frac{1}{2\beta^2}$. By specifying the two parameters $b$ and $\gamma$, the original form of the EH Lagrangian (\ref{EH Lagrangian}) up to the leading order of corrections is simply recovered. The Lagrangian density (\ref{EH}) can be considered as the simplest model of BI-type $U(1)$ gauge theories which matches with the weak-field limit of BI-type and EH theories. In order to analyze the self-energy problem in section \ref{sec:self-energy}, this Lagrangian will be considered as a generic Lagrangian and it does not matter whether we work in weak- or strong-field limits (the reason will be clarified in section \ref{sec:3.3}). Besides the Maxwell Lagrangian, a quadratic Maxwell invariant term also appears in the low energy limit of $E_8 \times E_8$ heterotic string theory \cite{Gross1987}, which gives the study of the Lagrangian (\ref{EH}) in higher dimensions a special significance.

BI-type theories we have shown in eq. (\ref{BI-type theories}) were originally developed and studied in \cite{Born-Infeld1934,Soleng,Hendi2012} case by case. But, they can be studied in a more general treatment assuming that the asymptotic behavior of them is the same as eq. (\ref{asymp-BI-EH}), as will be evident in the next sections. For later applications, we also introduce the derivative of BI-type Lagrangians with respect to the invariant $\cal S$ in the weak-field coupling limit, which takes the following form

\begin{equation}
{\left. {\frac{{\partial {\cal L}({\cal S})}}{{\partial {\cal S}}}} \right|_{{\rm{large}}\,\beta }} =  - 1 + {c_1}\frac{{\cal S}}{{{\beta ^2}}} - {c_2}\frac{{{{\cal S}^2}}}{{{\beta ^4}}} + {c_3}\frac{{{{\cal S}^4}}}{{{\beta ^6}}} + O\left( {{\beta ^{ - 8}}} \right),
\end{equation}
in which, for the coefficient $c_1$, we can always set $c_1=1$ for any BI-type theory.

\subsection{The field equations} \label{sec:field eqs}

Varying the action of any model of BI-type $U(1)$ gauge theories with respect to the gauge field $A_\mu$ , the nonlinear electromagnetic field equations are found to be as

\begin{equation} \label{field eqs general}
{\nabla _\mu }\left( {\frac{{\partial {\cal L}}}{{\partial {\cal F}}}{F^{\mu \nu }} + \frac{{\partial {\cal L}}}{{\partial {\cal G}}}{{}^*{F}^{\mu \nu }}} \right) = 0.
\end{equation}
From this, one can straightforwardly obtain the following nonlinear Maxwell's equations
\begin{equation}
\nabla .{\bf{D}} = 0,\,\nabla  \times {\bf{H}} = \frac{{\partial {\bf{D}}}}{{\partial t}},
\end{equation}
where $\bf{D}$ and $\bf{H}$ are the electric displacement field and the magnetic field strength, respectively. Using the definition of the electric displacement field, i.e., ${\bf{D}} = \frac{{\partial {\cal L}}}{{\partial {\bf{E}}}}$ and the chain rule of differentiation as $\frac{{\partial {\cal L}}}{{\partial {\bf{E}}}}=\frac{{\partial {\cal L}}}{{\partial {\cal S}}}\frac{{\partial {\cal S}}}{{\partial {\bf{E}}}}$, one obtains
\begin{equation}\label{D field}
{\bf{D}} = \left( {{\bf{E}} + 2\frac{\gamma }{{{\beta ^2}}}\left( {{\bf{E}}.{\bf{B}}} \right){\bf{B}}} \right)\left( { - \frac{{\partial {\cal L}}}{{\partial {\cal S}}}} \right)
\end{equation}
The magnetic field strength is defined by $\bf{H}=-\frac{\partial {\bf{\cal{L}}}}{\partial {\bf{B}}}$, yielding 

\begin{equation}\label{H field}
{\bf{H}} = \left( {{\bf{B}} - 2\frac{\gamma }{{{\beta ^2}}}\left( {{\bf{E}}.{\bf{B}}} \right){\bf{E}}} \right)\left( { - \frac{{\partial {\cal L}}}{{\partial {\cal S}}}} \right)
\end{equation}
in which the chain rule as $\frac{{\partial {\cal L}}}{{\partial {\bf{B}}}} = \frac{{\partial {\cal L}}}{{\partial {\cal S}}}\frac{{\partial {\cal S}}}{{\partial {\bf{B}}}}$ has been used. Obviously, in the context of nonlinear electrodynamics, the fields $\bf{D}$ and $\bf{H}$ even in vacuum are no longer simply multiples of $\bf{E}$ and $\bf{B}$. The rest of nonlinear Maxwell's equations can be found by use of the Bianchi identity, ${\nabla _\mu }{{}^*F^{\mu \nu }} = 0$, as
\begin{equation}
\nabla .{\bf{B}} = 0,\,\nabla  \times {\bf{E}} =  - \frac{{\partial {\bf{B}}}}{{\partial t}}.
\end{equation}
As seen, these equations are the same as before in Maxwell's electrodynamics since they do not depend on the theory's Lagrangian.

\subsection{On the electrostatic field of elementary point charges}

Here we investigate the behavior of electrostatic field in BI-type theories in order to gain insight into the weak- and strong-field limits of these theories. This is essential for later applications in section \ref{sec:self-energy}. Our starting point is the field equation (\ref{field eqs general}), which in electrostatic considerations reduces to 
\begin{equation}
{\nabla _\mu }\left( {\frac{{\partial {\cal L}}}{{\partial {\cal F}}}{F^{\mu \nu }}} \right) = 0\,\,\, \to \,\,\,{\partial _\mu }\left( {\sqrt { - g} \frac{{\partial {\cal L}}}{{\partial {\cal F}}}{F^{\mu \nu }}} \right) = 0.
\end{equation}
$\sqrt { - g}$ is the determinant of the spacetime metric. Regarding the Minkowski background in spherical coordinates, the following simple equation for the electrostatic field of a point charge is obtained which is valid for all models of BI-type theories 
\begin{equation} \label{simple field equation}
{r^{d - 2}}\left( {\frac{{\partial {\cal L}}}{{\partial {\cal F}}}} \right)E(r) = C,\,\,\,\,\,\,\,({\cal F} =  - {{\bf{E}}^2}/2),
\end{equation}
Since in the weak-field limit we always have ${\frac{{\partial {\cal L}}}{{\partial {\cal F}}}} =  - 1 + O\left( {{\beta ^{ - 2}}} \right)$, the integration constant ($C$) is straightforwardly obtained as $C=-Q$, which is the usual Maxwell limit. To be more specific, the electrostatic fields in the models under consideration (in arbitrary dimensions) are obtained as\footnote{Using the property $W(x) \times {{\rm{e}}^{W(x)}} = x$, the electrostatic field in exponential model may be equivalently written as
\begin{equation}
{\bf{E}}(r) = \frac{Q}{{{r^{d - 2}}}}{{\rm{e}}^{ - \frac{1}{2}W\left( {\frac{{{Q^2}}}{{{\beta ^2}{r^{2(d - 2)}}}}} \right)}}\hat r
\end{equation}
which is more common in black hole physics  \cite{Hendi2012,Hendi2013,Hendi2015Dehghani}.}
\begin{equation} \label{electric fields}
{\bf{E}}{\rm{(r):}}\left\{ \begin{array}{l}
{{\bf{E}}_{{\rm{BI}}}}(r) = \frac{{\beta Q}}{{\sqrt {{Q^2} + {\beta ^2}{r^{2(d - 2)}}} }}\hat r\\
{{\bf{E}}_{\log }}(r) = \frac{{2Q}}{{{r^{d - 2}} + \sqrt {{r^{2(d - 2)}} + \frac{{2{Q^2}}}{{{\beta ^2}}}} }}\hat r\\
{{\bf{E}}_{\exp }}(r) = \beta \sqrt {W\left( {\frac{{{Q^2}}}{{{\beta ^2}{r^{2(d - 2)}}}}} \right)} \hat r\\
{{\bf{E}}_{{\rm{EH}}}}(r) = \left( {\sqrt[3]{{\frac{Q}{{2b{r^{d - 2}}}} + \sqrt {{{\left( {\frac{Q}{{2b{r^{d - 2}}}}} \right)}^2} + {{\left( {\frac{1}{{3b}}} \right)}^3}} }} - \sqrt[3]{{ - \frac{Q}{{2b{r^{d - 2}}}} + \sqrt {{{\left( {\frac{Q}{{2b{r^{d - 2}}}}} \right)}^2} + {{\left( {\frac{1}{{3b}}} \right)}^3}} }}} \right)\hat r
\end{array} \right.
\end{equation}
where $W(x)$ is the Lambert $W$ function (also called the product logarithm or less commonly the omega function). In the weak-field coupling limit $\beta \to \infty$ (not necessarily infinite, but large enough), the expansion of the electric field for any BI-type action is of the form 
\begin{equation} \label{E-field-weak}
{\left. {E(r)} \right|_{{\rm{large}}\,\beta }} = \frac{q}{{{r^{d - 2}}}} - \frac{{{q^3}}}{{2{\beta ^2}{r^{2(d - 2)}}}} + O\left( {{\beta ^{ - 4}}} \right).
\end{equation}
Note that, generally in BI-type theories, the weak-field coupling limit ($\beta \to \infty$) is necessarily equivalent to the expansion around $r=\infty$. This key feature is implied from the fact we have already demanded that the nonlinear theory should be turned into Maxwell's theory in the weak-field limit, i.e., eq. (\ref{asymp-BI-EH}).

\begin{figure}[!htbp]
	\begin{center}
		\epsfxsize=10 cm 
		\includegraphics[width=11 cm]{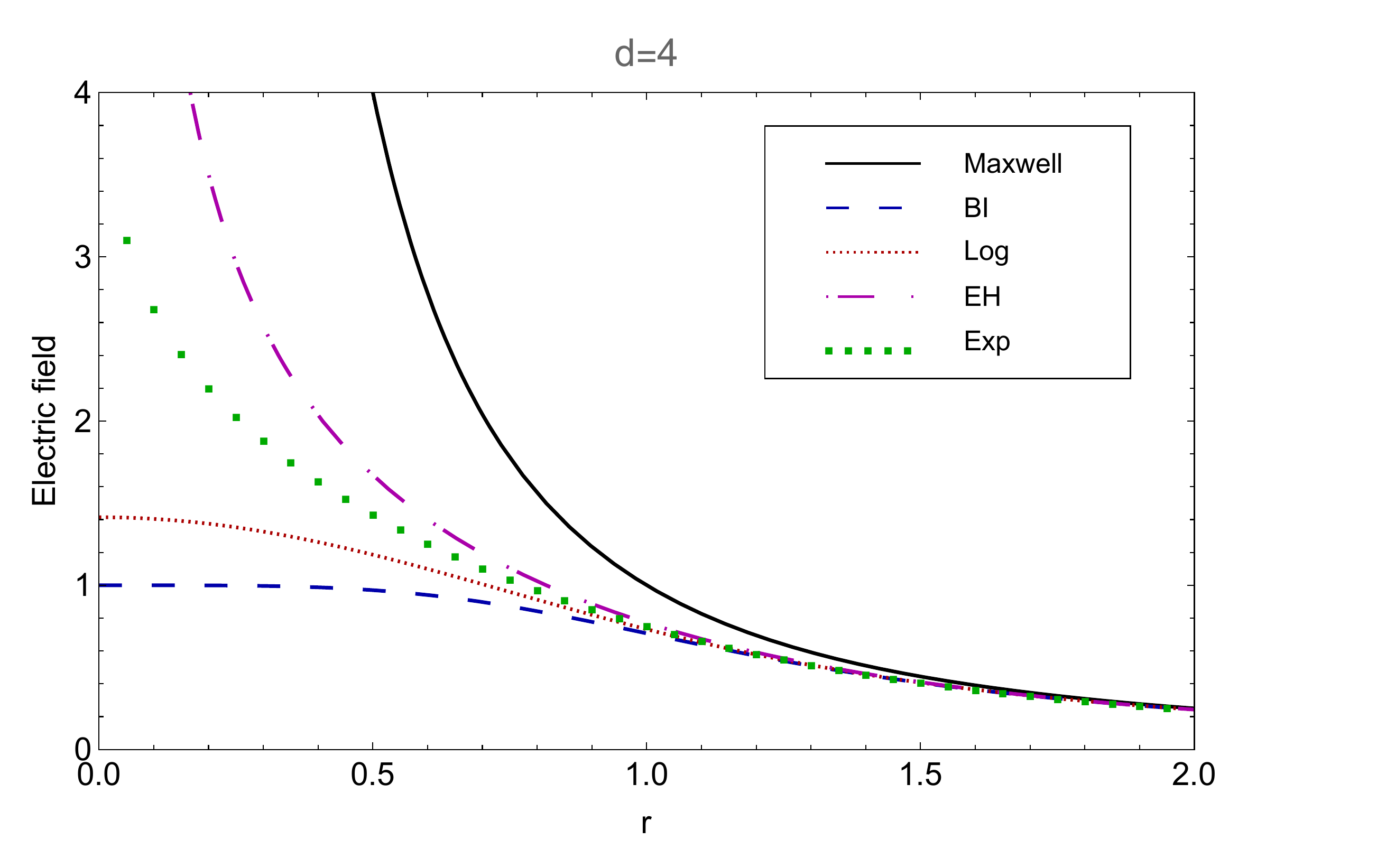}
		\caption{The behavior of the electric field near a point charge in the Born-Infeld (BI), logarithmic, exponential and Euler-Heisenberg (EH) theories in $d=4$ dimensions. We have set $Q=1$ and $\beta = 1$. Note that in the case of EH model the parameter $b$ has been already identified as $b = \frac{1}{2\beta^2}$.}
		\label{electric field_d4_all}
	\end{center}
\end{figure}

However, BI-type gauge theories can behave differently in strong-field limit and the outcomes in this limit should be examined for each theory separately. But we can draw general conclusions as well. Now, let us consider the strong-field limit which is essentially equivalent to the expansions around $\beta = 0$ or $r = 0$. In the strong-field limit, for the models under consideration we explicitly have
\begin{equation} \label{E-field-strong}
{\bf{E}}{\rm{(r):}}\left\{ \begin{array}{l}
{E_{{\rm{BI}}}}(r) = \beta  - \frac{{{\beta ^3}}}{{2{Q^2}}}{r^{2(d - 2)}} + O\left( {{\beta ^5}} \right)\\
{E_{\log }}(r) = \sqrt 2 \beta  - \frac{{{\beta ^2}}}{Q}{r^{d - 2}} + O\left( {{\beta ^3}} \right)\\
{E_{\exp }}(r) = \beta \sqrt {\ln \left( {\frac{{{Q^2}}}{{{\beta ^2}{r^{2(d - 2)}}}}} \right) - \ln \left( {\ln \left( {\frac{{{Q^2}}}{{{\beta ^2}{r^{2(d - 2)}}}}} \right)} \right)} \\
{E_{{\rm{EH}}}}(r) = \frac{1}{{{2^{2/3}}}}{\left( {\frac{Q}{b}} \right)^{1/3}}{r^{ - \left( {\frac{{d - 2}}{3}} \right)}} - \frac{1}{{6 \times {2^{1/3}}}}{\left( {\frac{Q}{b}} \right)^{2/3}}{r^{\left( {\frac{{d - 2}}{3}} \right)}} + O\left( {{r^{\frac{{5(d - 2)}}{3}}}} \right)
\end{array} \right.
\end{equation}
These Taylor series about $r=0$ are very interesting and are helpful for analyzing the self-energy problem in BI-type theories, as will be evident in section \ref{sec:self-energy} (Fig. \ref{electric field_d4_all}).

\section{Finiteness of the field energy of elementary point charges} \label{sec:self-energy}

In this section, we first concentrate on the self-energy problem in exponential $U(1)$ gauge theory. Usually, when the electric field is singular at the origin (where the point charge is located), it is considered as a sign of infinite self-energy. Having an upper limit on the electric field of a point charge trivially leads to a finite self-energy. It is shown that it is not mandatory to have an upper limit on the electric field in order for the self-energy of point charges to be finite. In addition to EH theory \cite{Shabad2015} having the quadratic Lagrangian density (\ref{EH}), this is the second example of such theories that behaves in this way. Then we bring some new perspectives on this issue in higher dimensions for BI family of nonlinear $U(1)$ gauge theories.

\subsection{The self-energy issue in exponential nonlinear electrodynamics}

Here, we will show that the the total energy stored in the electrostatic field of a point charge in exponential nonlinear electrodynamics is finite, albeit the electric field at the center of a charge is still infinite. This feature has already been seen in the EH model of nonlinear electrodynamics \cite{Shabad2015}, but so far such behavior has not been observed in any other BI-type model. 

Now, let us focus our attention on the case of exponential model. Generally, the stress-energy tensor for the electromagnetic field in the cgs units is defined as 
\begin{equation}\label{set}
T_\nu ^\mu  =\frac{1}{4\pi}\Bigg[ \frac{{\partial {\cal L}}}{{\partial {\cal F}}}{F^{\mu \rho }}{F_{\nu \rho }} + \frac{{\partial {\cal L}}}{{\partial {\cal G}}}{{}^* F^{\mu \rho }}{F_{\nu \rho }} - \delta _\nu ^\mu {\cal L}\Bigg].
\end{equation}
For the case of exponential nonlinear electrodynamics, the energy density ($\rho_E$) which is the $T_0 ^0$ component reads
\begin{equation}
\rho_E \equiv T_0^0 = \frac{1}{4\pi}\Bigg[ - {{\mathop{\rm e}\nolimits} ^{ - {\cal S}/{\beta ^2}}}{F^{0\rho }}{F_{0\rho }} + \frac{1}{{{\beta ^2}}}{\cal G}{{\mathop{\rm e}\nolimits} ^{ - {\cal S}/{\beta ^2}}}{{}^* F^{0\rho }}{F_{0\rho }} - {\beta ^2}\left( {{{\mathop{\rm e}\nolimits} ^{ - {\cal S}/{\beta ^2}}} - 1} \right)\Bigg].
\end{equation}
Assuming ${\bf{B}}=0$, the electrostatic energy density is given by
\begin{equation}\label{rho_exp}
{\rho _E} = \frac{1}{4\pi}\Big(\left( {{{\bf{E}}^2} - {\beta ^2}} \right){{\mathop{\rm e}\nolimits} ^{ - {\cal S}/{\beta ^2}}} + {\beta ^2\Big)}.
\end{equation}
Making use of eq. (\ref{electric fields}) for $d=4$ and eq. (\ref{rho_exp}), finally, the self-energy of a point-like charge in the exponential nonlinear electrodynamics reads  

\begin{eqnarray}
{\cal E}_{\exp} &=& \int {{d^3}{\bf{x}}\sqrt { - g} } \,T_0^0\\ \nonumber
&=& 4\pi \int\limits_0^\infty  {{\rho _E}{r^2}dr} \\ \nonumber
&=&  \int\limits_0^\infty  {{r^2}\left( {{{\rm{e}}^{\frac{1}{2}{W}\left( {\frac{{{Q^2}}}{{{\beta ^2}{r^4}}}} \right)}}\left( {{\beta ^2}{W}\left( {\frac{{{Q^2}}}{{{\beta ^2}{r^4}}}} \right) - {\beta ^2}} \right) + {\beta ^2}} \right)} dr.
\end{eqnarray}
The above integral can be solved analytically by change of variables as
\begin{equation}
\frac{{{Q^2}}}{{{\beta ^2}{r^4}}} = x \to {r^2} = \frac{Q}{{\beta \sqrt x }}\,\,\,{\rm{and}}\,\,\,dr =  - \frac{{\sqrt Q }}{{4\sqrt \beta  {x^{5/4}}}}.
\end{equation}
It is a matter of calculation to show that the self-energy of a point charge like electron in exponential nonlinear electrodynamics is calculated as
\begin{eqnarray} \label{se_exp}
{{\cal E}_{\exp }} &=&\left. {\frac{1}{3} {\beta ^2}{r^3} - \frac{{\sqrt \beta  {Q^{3/2}}}}{{8\sqrt 2 }}\left( {\Gamma \left( { - \frac{3}{4},\frac{1}{4}W\left( {\frac{{{Q^2}}}{{{\beta ^2}{r^4}}}} \right)} \right) - 16\,\Gamma \left( {\frac{5}{4},\frac{1}{4}W\left( {\frac{{{Q^2}}}{{{\beta ^2}{r^4}}}} \right)} \right)} \right)} \right|_0^\infty \nonumber \\ 
&=&  - \frac{{\sqrt \beta   {Q^{3/2}}\left( {\Gamma \left[ { - \frac{3}{4}} \right] - 16\,\Gamma \left[ {\frac{5}{4}} \right]} \right)}}{{8\sqrt 2 }} \nonumber \\
&=& 1.709\sqrt {{Q^3}\beta }.
\end{eqnarray}

\begin{figure}[!htbp]
	\begin{center}
		\epsfxsize=10 cm 
		\includegraphics[width=7.2 cm]{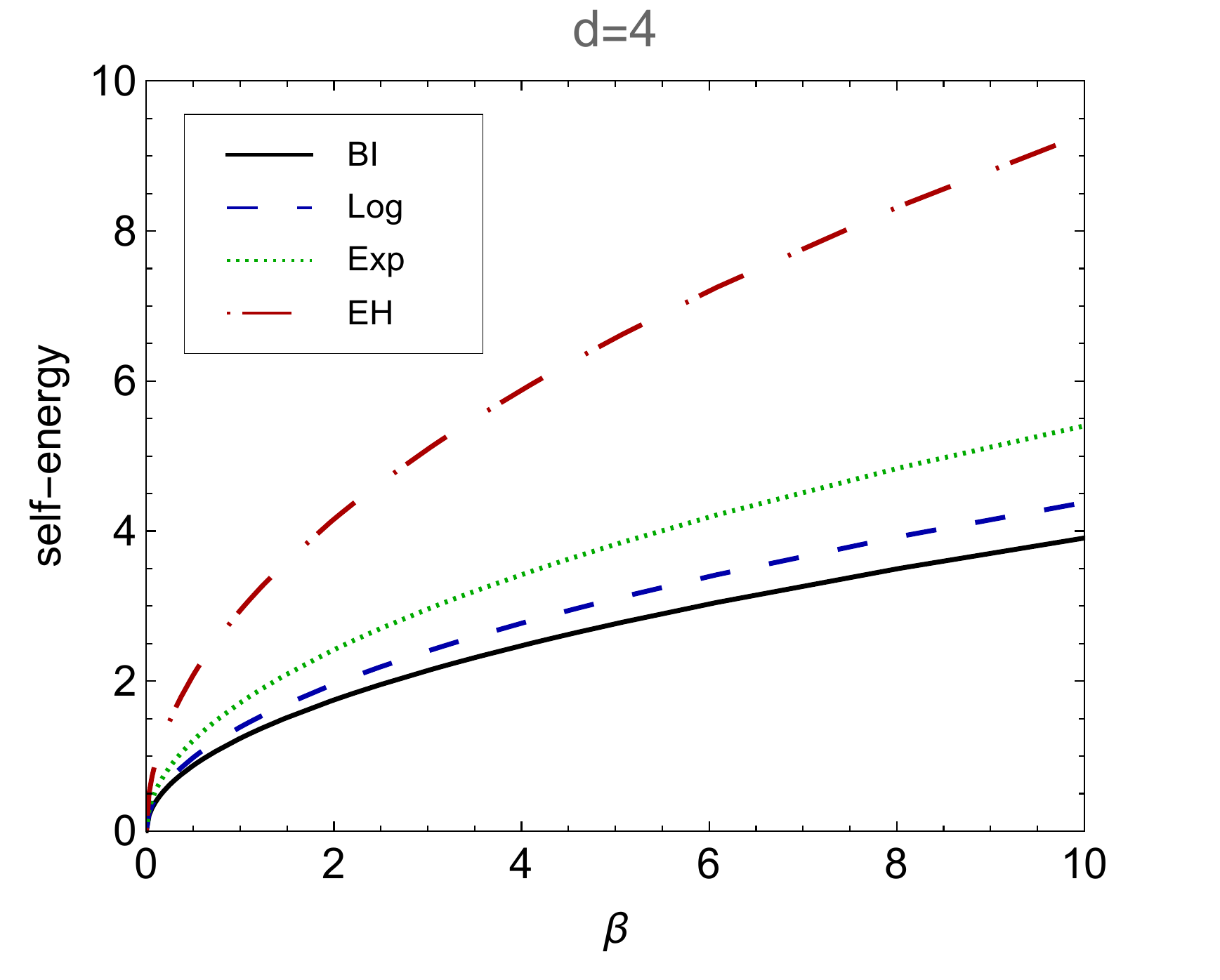}
		\hskip 0.1 cm
		\includegraphics[width=7 cm]{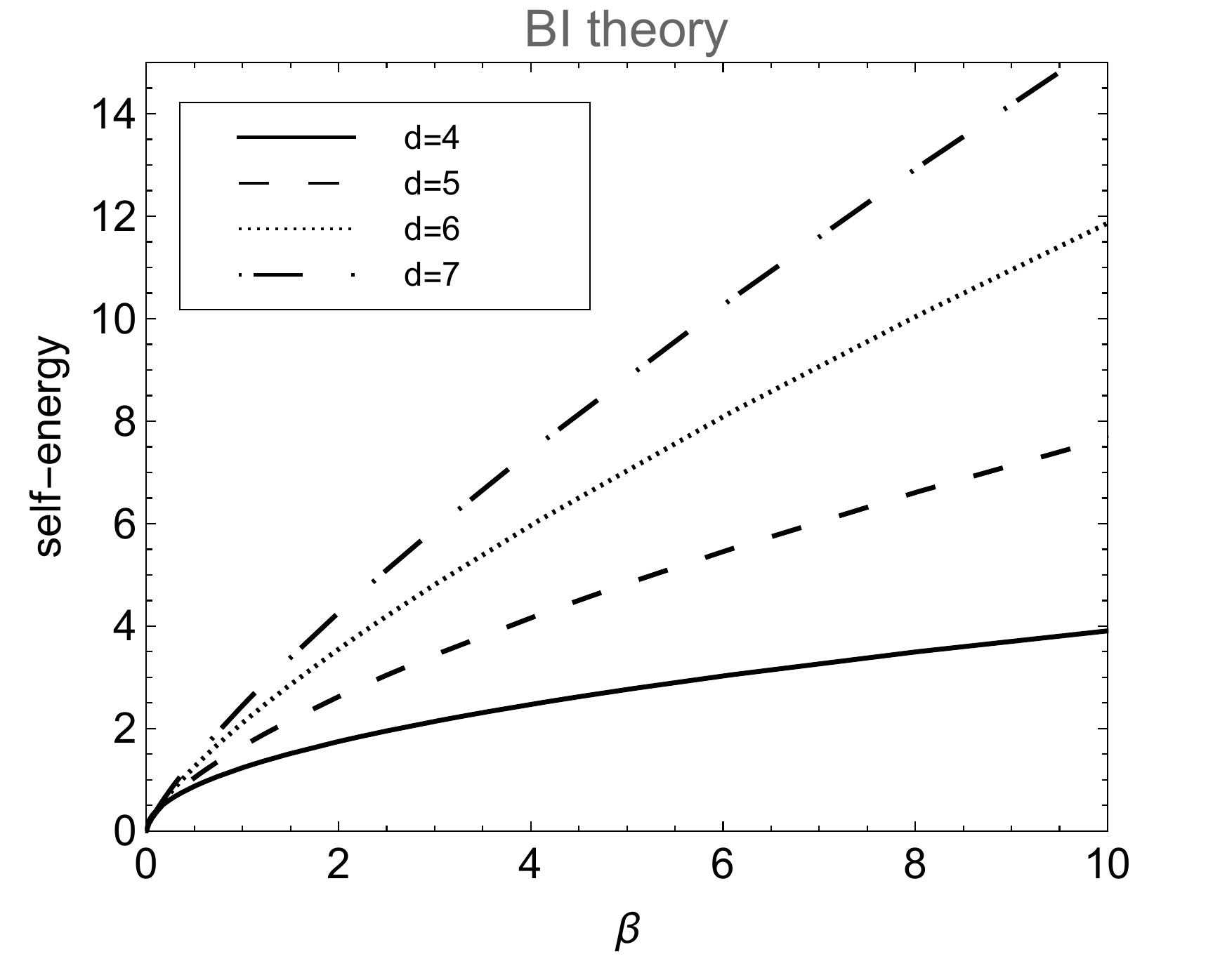}
		\caption{\textit{Left panel}: 	The self-energy of a point-like charge in the Born-Infeld (BI), logarithmic (Log), exponential (Exp) and Euler-Heisenberg (EH) theories in $d=4$ dimensions. \textit{Right panel}: 	The self-energy of a point-like charge in the BI nonlinear electrodynamics in various dimensions. This behavior is observed qualitatively for all BI-type models (Of course, if the self-energy in higher dimensions does not diverge for the model under consideration.) We have also set $Q=1$ for both figures.}
		\label{exp_self energy}
	\end{center}
\end{figure}

We see that contrary to Maxwell's electrodynamics, here the self-energy has a finite value despite the singularity of electric field (\ref{electric fields}) at $r=0$. This comes from the fact that the growth rate of the electrostatic field of exponential nonlinear electrodynamics is much slower than the Coulomb field as $r$ goes to zero. The left panel in figure \ref{exp_self energy} shows the self-energy of a point-like charge versus $\beta$ in different BI-type theories. As expected, in all cases, the self-energy increases as $\beta$ increases and eventually goes to infinity as $\beta  \to \infty $ (the Maxwell limit). 

By equating the self energy of the electron to its rest mass, $m_{e}c^2$, an order-of-magnitude estimate for the nonlinear parameter $\beta$ can be found. Note that in addition to electromagnetic interactions, leptons also participate in weak and gravitational processes and hence ${\cal E}=m_{e}c^2$ gives an upper limit for $\beta$. Equating  eq. (\ref{se_exp}) to the electron rest mass, the nonlinear parameter $\beta$ in cgs units is obtained
\begin{equation}
\beta _{\rm{exp}}= 2.070703876669462 \times {10^{15}}\,{\rm{e}}{\rm{.s}}{\rm{.u,}}\nonumber\\
\end{equation}
where
\begin{eqnarray}
{m_e} =&& 9.1093837 \times {10^{ - 28}}\,{\rm{g}},\nonumber\\
c =&& 2.99792458 \times {10^{10}}\,{\rm{cm/s}},\nonumber\\ 
e =&& 4.80320427 \times {10^{ - 10}}\,{\rm{cm}}^{\rm{3/2}}{{\rm{g}}^{\rm{1/2}}}{{\rm{s}}^{\rm{ -1 }}},
\end{eqnarray}
and in the SI units
\begin{equation} \label{exp beta}
\beta _{\rm{exp}} =\frac{k}{{10c \times {{10}^{ - 4}}}}\beta _{{\rm{esu}}} =6.207811775935659 \times {10^{19}}\,{\rm{V/m}},
\end{equation}
where $k = 8.9875517923 \times {10^9}$. The nonlinear parameter $\beta$ in BI, logarithmic and EH models which has already been obtained \cite{Born-Infeld1934,Gaete2014a,Shabad2015}, in the SI system of units are 

\begin{eqnarray} \label{theoretical beta}
{\beta _{{\rm{BI}}}} =&& 1.186905794319877 \times {10^{20}}\,{\rm{V/m}},\nonumber\\
{\beta _{{\rm{log}}}} =&& 9.441777777673394 \times {10^{19}}\,{\rm{V/m}},\nonumber\\
{\beta _{{\rm{EH}}}} = && 2.098172549250914 \times {10^{19}}\,{\rm{V/m}}.
\end{eqnarray}

Interestingly, in all cases, the values obtained for the nonlinear parameter are greater than the so-called Schwinger limit\footnote{Also known as the Sauter–Schwinger effect \cite{Sauter1931,Schwinger1951}.}, i.e., \cite{HeisenbergEuler1936,Sauter1931,Schwinger1951}
\begin{equation} \label{Schwinger limit}
{E_c} =c {B_c} = \frac{{{m^2_e}{c^3}}}{{e{\hbar}}} = 1.32 \times {10^{18}}\,{\rm{V/m}},
\end{equation}
where $E_c$ and $B_c$ are called critical fields. The main reason probably is that we have taken all the energy of the electrostatic field stored in space equal to the relativistic rest mass. But the important point is that even if the electromagnetic mass is part of the relativistic rest mass, then the nonlinear parameter $\beta$ will be comparable to the Schwinger scale. As discussed in a number of pioneering works in QED \cite{Sauter1931,HeisenbergEuler1936,Schwinger1951}, above this scale, the nonlinearity of electromagnetic fields becomes sensible and the effects of vacuum polarization become noticeable. Interestingly, in BI-type theories where the upper limit on the electric field do not exist, the value obtained for the nonlinear parameter $\beta$ is slightly closer to the Schwinger limit.

Therefore, it is possible not to set the upper limit on the electric field, but still obtain a finite value for the self-energy of point charges. In such BI-type models, the nonlinear parameter $\beta$ can still be fixed by calculating the self-energy, which is generally comparable to the Schwinger limit (\ref{Schwinger limit}). In these kinds of BI-type models, the electric field grows larger and larger as we get closer to the point charge in agreement with the current empirical evidence\footnote{The electromagnetic fields as large as $10^{25} \rm{V}/\rm{m}$ has been detected by ATLAS \cite{ATLAS2017}.}, but the total field energy remains finite, as we expect from our physical intuition.

\subsection{The self-energy issue in higher dimensional BI-type theories}

Before proceeding, we would like to clarify about some points in this and the next subsections. The BI-type Lagrangians have been found and studied in higher dimensions with different motivations and the Taylor expansion of these Lagrangians in the weak-field limit takes the form of eq. (\ref{asymp-BI-EH}). Since in four-dimensions this weak-field limit is structurally the same as the weak-field limit of EH theory (\ref{EH Lagrangian}), we use the same terminology in higher dimensions and refer to it as (weak-field) EH Lagrangian. We emphasize that it is true that the mathematical form of EH Lagrangian is the same as the weak-field limit of BI-type models, but it does not correspond to a one-loop effective action of QED in higher dimensions (for a discussion about the form of the one-loop Lagrangians in higher-dimensional QED see ref. \cite{Ritz1996}). However, all of these (BI-type and weak-field EH) theories are in strong connection with string theory \cite{StringBI1,StringBI2,Gross1987,StringBI3,StringBI4}.  On the other hand, the weak-field EH Lagrangian (\ref{EH}), which is considered as a generic Lagrangian in this section, will play a vital role in order to derive some novel conclusions about the self-energy problem in BI-type family of $U(1)$ gauge theories in arbitrary dimensions, as will be clarified in the next subsection \ref{sec:3.3}.

It is not clear whether the the self-energy of elementary point charges in higher dimensions are finite or not. Here, we explicitly confirm the finiteness of the self-energy in higher dimensions for exponential nonlinear electrodynamics as well as for the cases of BI and logarithmic models of nonlinear electrodynamics. Furthermore, we will prove that unlike the finiteness of self-energy in EH theory in 4-dimensional spacetime \cite{Shabad2015}, it diverges in higher dimensions. Since EH theory has been identified as the weak-field limit of BI-type theories, the latter finding will lead us to interesting results, as will be discussed further in the next section.

Our starting point is the definition of self-energy in $d$-dimensional spacetime, given by
\begin{equation}\label{self}
{\cal E} = \frac{-1}{{4\pi }}\int\limits_0^\infty  {\left( { {{\bf{E}}^2}\frac{{\partial {\cal L}}}{{\partial {\cal F}}} +{\cal L}} \right)\,} dV,
\end{equation}
where $dV$ is the $d$-dimensional volume element in spherical coordinates 
\begin{equation}
dV = \frac{{\left( {d - 1} \right){\pi ^{\frac{{d - 1}}{2}}}}}{{\Gamma \left( {\frac{{d - 1}}{2} + 1} \right)}}{r^{d - 2}}dr.
\end{equation}
For the self-energy of a point charge in exponential nonlinear electrodynamics, the explicit computation shows that the self energy in higher dimensions is always finite and can be found from the following general relation
\begin{eqnarray}
{{\cal E} _{\exp }} = &&\frac{1}{{\Gamma \left( {\frac{{d + 1}}{2}} \right)}}{2^{\frac{{7d + 7}}{{4 - 2d}}}}{\left( {d - 2} \right)^{\frac{{5 - 3d}}{{2\left( {d - 2} \right)}}}}{\pi ^{\frac{{d - 3}}{2}}}{\beta ^{\frac{{d - 3}}{{d - 2}}}}{Q^{\frac{{d - 1}}{{d - 2}}}} \times\nonumber \\
&&\left( {{2^{\frac{{2d}}{{d - 2}}}}\left( {{2^{\frac{6}{{d - 2}}}}{d^3} - {{52}^{\frac{6}{{d - 2}}}}{d^2} + {2^{\frac{{3d}}{{d - 2}}}}d - {2^{\frac{{2\left( {d + 1} \right)}}{{d - 2}}}}} \right)\Gamma \left( {\frac{{7 - 3d}}{{4 - 2d}}} \right) - {2^{\frac{{10}}{{d - 2}}}}\left( {d - 1} \right)\Gamma \left( {\frac{{1 - d}}{{2\left( {d - 2} \right)}}} \right)} \right)\nonumber\\
\end{eqnarray}
The numerical values of the self-energy of a point charged particle with charge $Q$ in 4- and higher dimensions in exponential nonlinear electrodynamics are given in the last row of table \ref{tab:self-energy}.

Now, we turn our attention to the cases of BI and logarithmic models. By generalizing the calculations done in ref. \cite{Born-Infeld1934} to higher dimensions, the self-energy of a point charge in BI nonlinear electrodynamics in $d$-dimensional space time is given by
\begin{equation}
{{\cal E} _{{\rm{BI}}}} =  - \frac{{{\pi ^{\frac{d}{2} - 2}}\Gamma \left( { - \frac{{d - 1}}{{2\left( {d - 2} \right)}}} \right)\Gamma \left( {1 + \frac{1}{{2d - 4}}} \right)}}{{4\Gamma \left( {\frac{{d - 1}}{2}} \right)}}{\beta ^{\frac{{d - 3}}{{d - 2}}}}{Q^{\frac{{d - 1}}{{d - 2}}}}.
\end{equation}
The numerical values of the self-energy in $d=4,5,6$ and $7$ in BI electrodynamics are given in the first row of table \ref{tab:self-energy}.

By extending the calculations done in ref. \cite{Gaete2014a} to higher dimensions, the general relation for the self-energy of a point charge in logarithmic nonlinear electrodynamics reads
\begin{equation}
{{\cal E} _{\log }} =  - \frac{{{2^{\frac{1}{2}\left( {\frac{1}{{d - 2}} - 3} \right) - 2}}{\pi ^{\frac{d}{2} - 2}}\Gamma \left( { - \frac{{d - 1}}{{2\left( {d - 2} \right)}}} \right)\Gamma \left( {\frac{1}{{2d - 4}}} \right)}}{{\Gamma \left( {\frac{{d + 1}}{2}} \right)}}{\beta ^{\frac{{d - 3}}{{d - 2}}}}{Q^{\frac{{d - 1}}{{d - 2}}}}
\end{equation}
The numerical values of the self-energy in logarithmic nonlinear electrodynamics in $4,5,6$ and $7-$ dimensional space time are given in the third row of table \ref{tab:self-energy}. Note that all the numerical values printed in this table are found in the cgs units and this is why the self-energy value we have given for $d=4$ in logarithmic nonlinear electrodynamics differs from eq. (33) of ref. \cite{Gaete2014a} in which the authors worked in the Lorentz-Heaviside units. To convert the cgs value for the self-energy to its corresponding value in the Lorentz-Heaviside system of units, we should multiply the cgs value by $\frac{4 \pi}{(4 \pi)^{3/2}}$.\footnote{The $4\pi$ factor in the numerator denotes the fact that in the Lorentz-Heaviside units where $\epsilon_{0}=\mu_{0}=1$, the stress-energy tensor (\ref{set}) does not possess $\frac{1}{4 \pi}$ factor. The $(4 \pi)^{3/2}$ factor in the denominator comes from the fact that in the cgs units ${\bf{D}} = \frac{e}{r^2}\hat r$ but in the Lorentz-Heaviside units ${\bf{D}} = \frac{e}{4\pi r^2}\hat r$.}

\begin{table}[]
	\centering
	\caption{The exact values of the self-energy for elementary point particles with charge $Q$ in $4$- and higher dimensions}
	\label{tab:self-energy}
	\begin{tabular}{|c|c|c|c|c|}
		\hline
		\textit{Theory}                                                                 & \textit{\begin{tabular}[c]{@{}c@{}}Self-energy\\ in d=4\end{tabular}} & \textit{\begin{tabular}[c]{@{}c@{}}Self-energy\\ in d=5\end{tabular}} & \textit{\begin{tabular}[c]{@{}c@{}}Self-energy\\ in d=6\end{tabular}} & \textit{\begin{tabular}[c]{@{}c@{}}Self-energy\\ in d=7\end{tabular}} \\ \hline
		\begin{tabular}[c]{@{}c@{}}Born-Infeld\\ electrodynamics\end{tabular}          & $1.236\,{\beta ^{\frac{1}{2}}}{Q^{\frac{3}{2}}}$ & $1.652\,{\beta^{\frac{2}{3}}}{Q^{\frac{4}{3}}}$                              & $2.110\,{\beta^{\frac{3}{4}}}{Q^{\frac{5}{4}}}$                              & $2.448\,{\beta^{\frac{4}{5}}}{Q^{\frac{6}{5}}}$                                \\ \hline
		\begin{tabular}[c]{@{}c@{}}Euler-Heisenberg\\ electrodynamics\end{tabular}
		& $2.940\,{\beta ^{\frac{1}{2}}}{Q^{\frac{3}{2}}}$ & infinite
		& infinite
		& infinite    \\ \hline
		\begin{tabular}[c]{@{}c@{}}Logarithmic nonlinear\\ electrodynamics\end{tabular} &$1.386\,{\beta ^{\frac{1}{2}}}{Q^{\frac{3}{2}}}$ &$1.967\,{\beta ^{\frac{2}{3}}}{Q^{\frac{4}{3}}}$ & $2.603\,{\beta ^{\frac{3}{4}}}{Q^{\frac{5}{4}}}$ & $3.092\,{\beta ^{\frac{4}{5}}}{Q^{\frac{6}{5}}}$ \\ \hline
		\begin{tabular}[c]{@{}c@{}}Exponential nonlinear\\ electrodynamics\end{tabular} &
		$1.709\,{\beta ^{\frac{1}{2}}}{Q^{\frac{3}{2}}}$ & $2.867\,{\beta ^{\frac{2}{3}}}{Q^{\frac{4}{3}}}$ & $4.331\,{\beta ^{\frac{3}{4}}}{Q^{\frac{5}{4}}}$ & $5.728\,{\beta ^{\frac{4}{5}}}{Q^{\frac{6}{5}}}$ \\ \hline
	\end{tabular}
\end{table}

We see that for BI, exponential and logarithmic nonlinear electrodynamics the self-energy in higher dimensions is finite as well as in $d=4$. However, this is not the case for EH theory. It was found that within this theory the self-energy is finite in $d=4$ dimensions \cite{Shabad2015}, nevertheless, here we will show that the EH theory may not resolve the self-energy issue in higher dimensions.

Using eq. (\ref{EH}) together with the self-energy definition in (\ref{self}), the self-energy of a point charge in EH electrodynamics reads
\begin{equation}\label{self_EH}
{{\cal E}_{{\rm{EH}}}} =\frac{1}{4\pi} \int\limits_0^\infty  {\left( {\frac{{{{\bf{E}}_{{\rm{EH}}}}^2}}{2} + \frac{{3b}}{4}{{\bf{E}}_{{\rm{EH}}}}^4} \right)dV} 
\end{equation}
Therefore we should calculate two integrals to determine the self-energy. Replacing the last row of (\ref{electric fields}) in eq. (\ref{self_EH}) and after two successive changing of variables\footnote{The first one is ${\left( {\frac{q}{{2b}}} \right)^{\frac{1}{{d - 2}}}}{x^{ - 1}}$ and the second one is $x = \left( {\frac{1}{{27{b^3}}}} \right){y^{ - 1}}$.}, the first integral in eq. (\ref{self_EH}) becomes
\begin{eqnarray}
{I_1} = &&\frac{{{2^{ - 4 + \frac{1}{{2 - d}}}}{3^{\frac{{d + 1}}{{2\left( {d - 2} \right)}}}}\left( {d - 1} \right){\pi ^{\frac{1}{2}\left( {d - 3} \right)}}}}{{\Gamma \left( {\frac{{d + 1}}{2}} \right)}}{b^{ - \frac{{d - 3}}{{2\left( {d - 2} \right)}}}}{q^{\frac{{d - 1}}{{d - 2}}}} \times  \nonumber\\
&&\int\limits_0^\infty  {{y^{\frac{{d - 2}}{3}}}{{\left( {{{\left( { - 1 + \sqrt {1 + {y^{2(d - 2)}}} } \right)}^{1/3}} - {{\left( {1 + \sqrt {1 + {y^{2(d - 2)}}} } \right)}^{1/3}}} \right)}^2}} dy.
\end{eqnarray}
This integral can be computed numerically and it converges for $d \ge 4$. Using changing of variables similar to what we did in computing $I_1$, the second integral in eq. (\ref{self_EH}) takes the following general form in $d$-dimensions
\begin{eqnarray}\label{I2}
{I_2} =&&\frac{{{2^{ - 5 + \frac{1}{{2 - d}}}}{3^{\frac{{d + 1}}{{2\left( {d - 2} \right)}}}}\left( {d - 1} \right){\pi ^{\frac{1}{2}\left( {d - 3} \right)}}}}{{\Gamma \left( {\frac{{d + 1}}{2}} \right)}}{b^{ - \frac{{d - 3}}{{2\left( {d - 2} \right)}}}}{q^{\frac{{d - 1}}{{d - 2}}}} \times \nonumber\\
&&\int\limits_0^\infty  {{y^{ - \frac{{d - 2}}{3}}}{{\left( { - {{\left( { - 1 + \sqrt {1 + {y^{2\left( {d - 2} \right)}}} } \right)}^{1/3}} + {{\left( {1 + \sqrt {1 + {y^{2\left( {d - 2} \right)}}} } \right)}^{1/3}}} \right)}^4}} dy.
\end{eqnarray}
The question of convergence of this integral ($I_2$) deserves further
investigation. The Taylor series of the integrand around $y=0$ is found to be as
\begin{equation} \label{Taylor series}
2 \times {2^{1/3}}{y^{ - \left( {\frac{{d - 2}}{3}} \right)}} - 4 \times {2^{2/3}}{y^{\frac{{d - 2}}{3}}} + O\left( {{y^{d - 2}}} \right).
\end{equation}
For $d=4$, this expansion equals to $\frac{{2 \times {2^{1/3}}}}{{{y^{2/3}}}} - 4 \times {2^{2/3}}{y^{2/3}} + O\left( {{y^{4/3}}} \right)$, so there is no singularity at $y=0$ and the integral converges. However, for $d>4$, it diverges and makes the self-energy of a point charge infinite in higher dimensional EH electrodynamics. For $d=5$, the expansion equals to $\frac{{2 \times {2^{1/3}}}}{y} - 4 \times {2^{2/3}}y + O\left( {{y^3}} \right)$ and the integral of $\frac{1}{y}$ is $\ln (y)$ which logarithmically diverges at $y=0$. In the same way it can be shown that $I_{2}$ is divergent for spacetime dimensions higher than five. As a result, from the finiteness of self-energy in a specific BI-type theory in one dimension, it can not be concluded that the self-energy remains finite in higher dimensions and this is a non-trivial result. \\

\subsection{Finiteness of the self-energy in BI-type family of $U(1)$ gauge theories} \label{sec:3.3}

\begin{figure}[!htbp]
	\begin{center}
		\epsfxsize=10 cm 
		\includegraphics[width=7 cm]{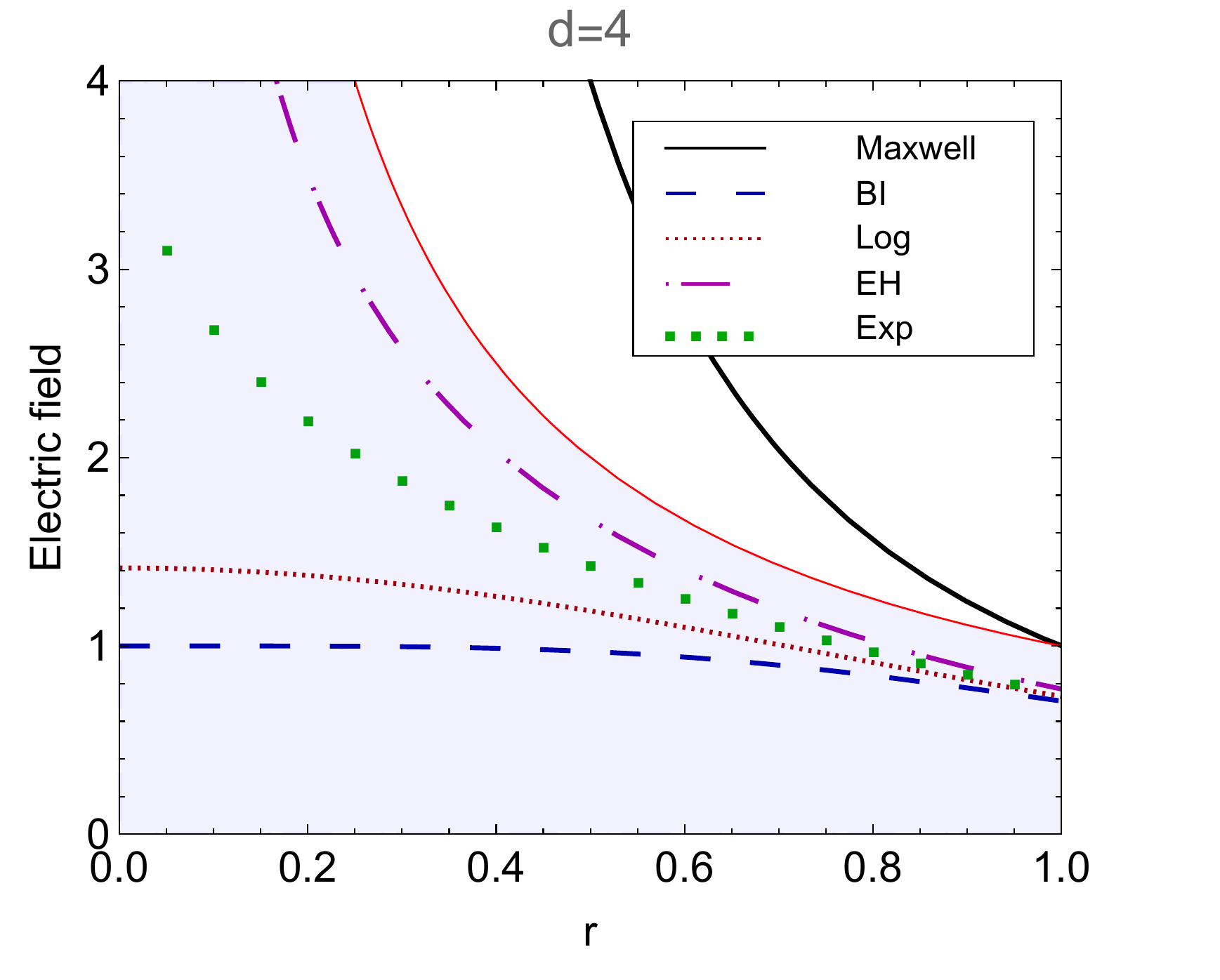}
		\hskip 0.1 cm
		\includegraphics[width=7 cm]{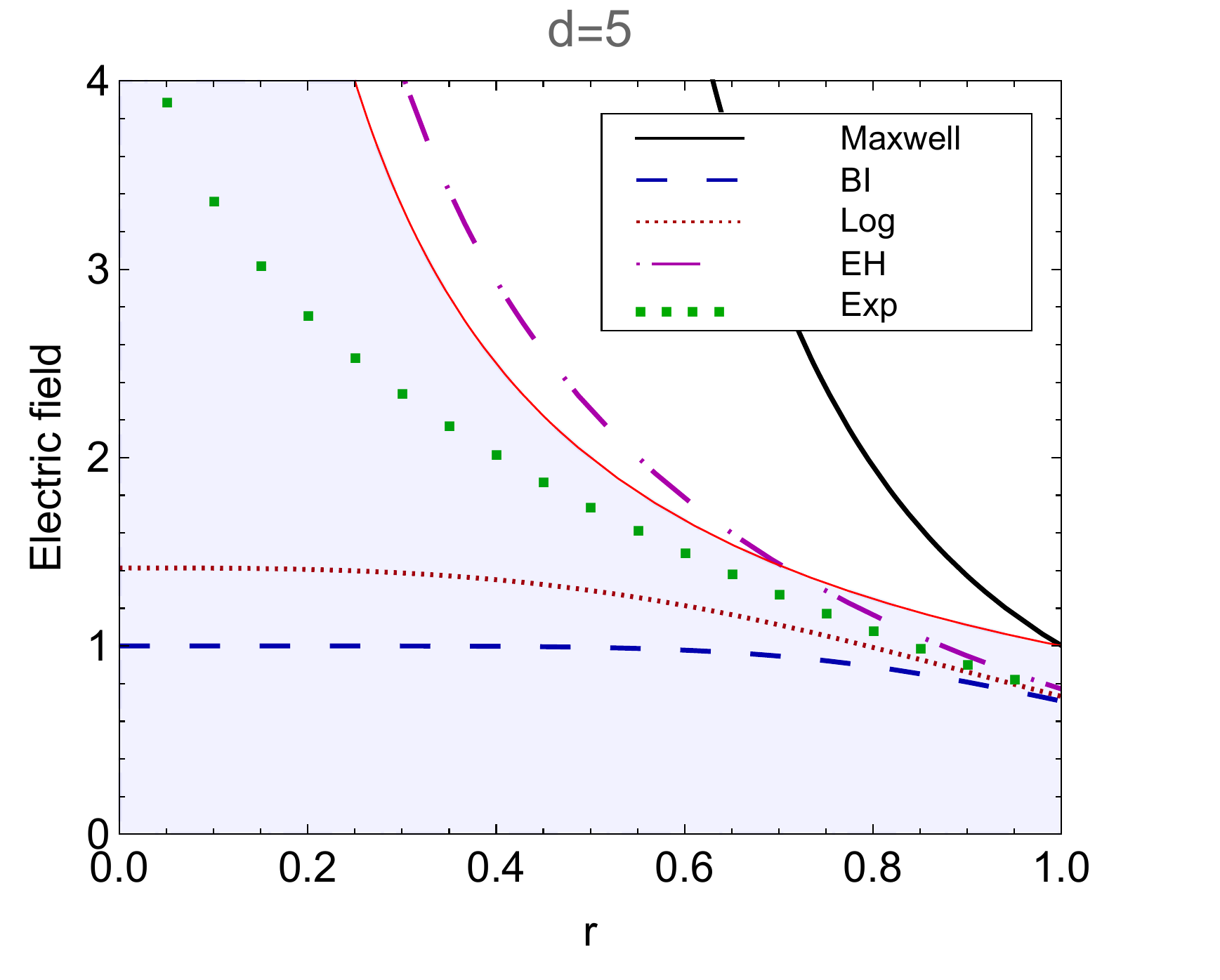}
		\vskip 0.1 cm
		\includegraphics[width=7 cm]{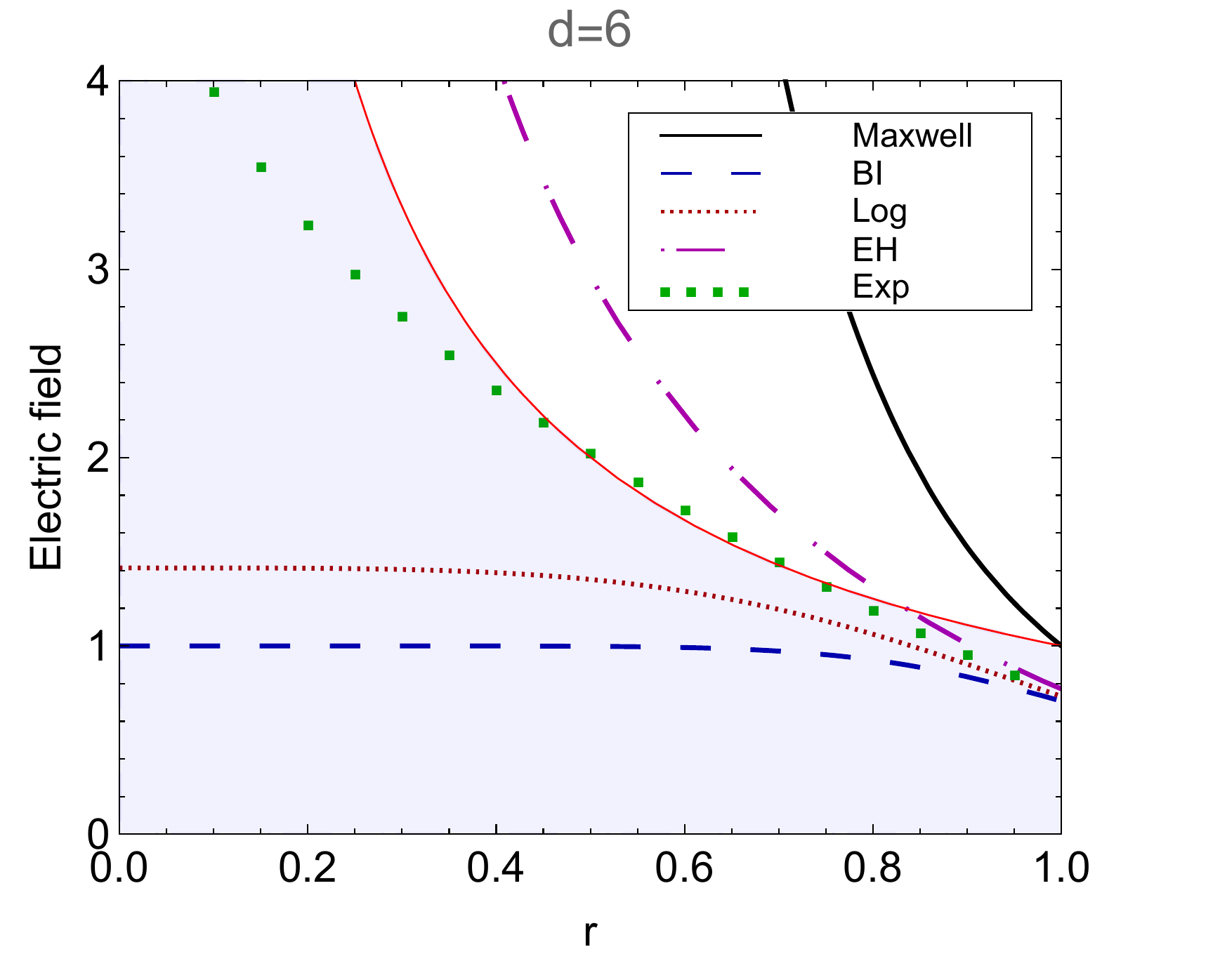}
		\hskip 0.1 cm
		\includegraphics[width=7 cm]{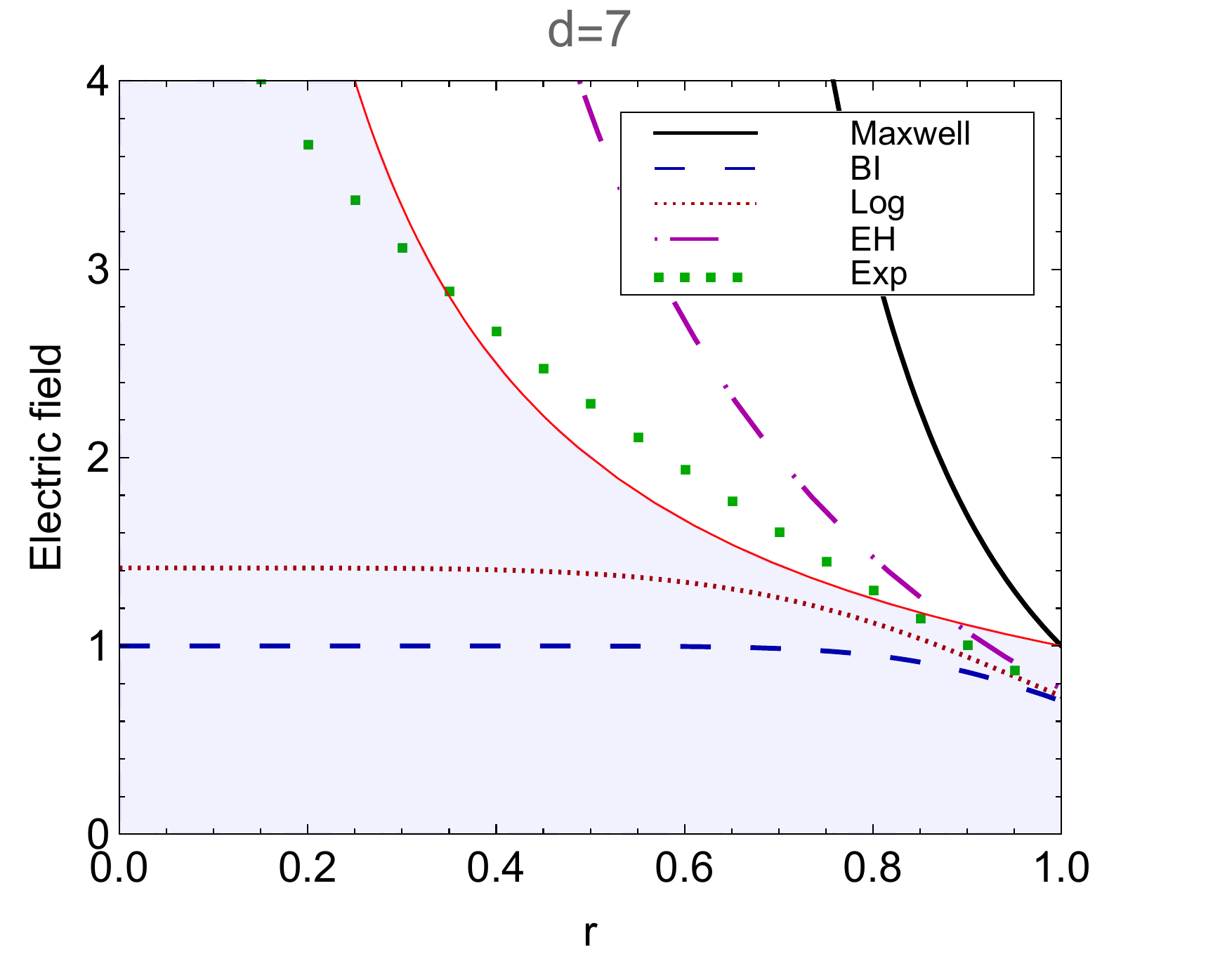}
		\caption{The behavior of the electric field in the Born-Infeld (BI), logarithmic, exponential and Euler-Heisenberg (EH) theories in $d=4,5,6$ and $7$ dimensions. The solid red lines represent the troublesome factor $\frac{1}{r}$.}
		\label{electric field_d4_7_all}
	\end{center}
\end{figure}

We can draw a number of important general conclusions from previous discussions. The effective EH theory with the Lagrangian density (\ref{EH Lagrangian}) is indeed the weak-field limit of BI-type gauge theories, some of them listed in relation (\ref{BI-type theories}). In the weak-field limit ($\beta \to \infty$ which is mathematically equivalent to $r \to \infty$), the theory is getting closer and closer to the outcomes of the Maxwell's electrodynamics theory; for instance, in this limit the electrostatic field of a point charge in any BI-type model is given by the expansion (\ref{E-field-weak}), in which the leading term is the so-called Coulomb's inverse-square law. This term was responsible for the infinity of self-energy in Maxwell's theory (it grows fast at the origin) but in BI-type theories it appears only as the first leading term of the Taylor expansion at large distances and obviously the Taylor expansion around $r = \infty$ is no longer valid as we get closer and closer to the electric charge. For BI-type theories, the electric fields are modified at distances very close to the electric charge in different ways. Some explicit examples have previously reported in eq. (\ref{E-field-strong}) for BI, logarithmic, exponential, and EH theories. In all cases, the electric field near the point charge grows slower than the corresponding Coulomb law. In figure \ref{electric field_d4_7_all}, the behavior of electric fields near the origin has been depicted for these theories in various dimensions.

Now, let us focus on the cases of 4-dimensional BI-type theories. Interestingly, in 4-dimensional spacetimes, the electric fields of BI-type theories grow slower than $\frac{1}{r}$ near the origin where the point charge is located.\footnote{In the cases of BI and logarithmic models, the electric fields rapidly reach the maximum field strength. In these cases, there does not exist any singularity at all.} The reason is that, in the EH theory as the weak-field limit of BI-type theories, the electric field at distances very close to the point charge behaves as $r^{-\frac{2}{3}}$ which grows much slower than $\frac{1}{r}$ (see figure \ref{electric field_d4_7_all}). So, for any other BI-type theory in 4-dimensions, the leading term of the Taylor expansion around $r=0$ grows slower than $\frac{1}{r}$. On the other hand, as shown in the previous section for the self-energy problem, the trouble (if exists any) is in the $r^{-\epsilon}$ factor in the Taylor expansion of integrand around $y=r=0$ (see eqs. (\ref{I2}) and (\ref{Taylor series})), in which ${\epsilon}$ is a positive real number. Obviously, the integral diverges when $\epsilon \ge 1$. In 4-dimensions, it always converges, meaning that divergences cannot appear at all. Therefore, we conjecture that, in 4-dimensions, the self-energy of elementary point charges is finite in any BI-type model of nonlinear $U(1)$ gauge theory.

This argument can straightforwardly extend to higher dimensional spacetimes as well. As proved in the previous section, the weak-field limit of higher-dimensional EH theory places neither an upper bound on the electric field, nor leads to a finite self-energy in higher dimensions ($d \ge 5$). As stated before, the reason is that the $\frac{1}{r}$ factor appears in the Taylor expansion of integrand (\ref{I2}) around $y=r=0$ and this leads to logarithmic divergences. Since the electric fields in other BI-type theories essentially grow slower than that of EH case with $\frac{1}{r}$, we conclude that BI-type theories in 5-dimensions always result in a finite self-energy and their weak-field limit, i.e., EH theory is an exception.\footnote{This discussion may be generalized to include other Lagrangians of nonlinear $U(1)$ electrodynamics that do not belong to BI-type family in order to see whether this claim is valid for them or not. In ref. \cite{Shabad2015}, the authors demonstrated that if the Lagrangian density of a nonlinear electrodynamics is any function of Maxwell invariant $\cal F$ which grows as a finite power of its argument, i.e., ${{\cal F}^{n + 1}}(n \ge 1)$ when ${\cal F} \to \infty$, then the integrand of the self-energy integral converges about $r=0$ as ${r^{\frac{{ - 2}}{{2n + 1}}}}$ and therefore the field energy of a charged point particle in $4$-dimensional spacetime will be finite. Note that the authors have not discussed the higher dimensional spacetimes. However, from the special case of $n=1$ which is the EH Lagrangian (\ref{EH}), it is evident that the criterion does not hold in higher dimensions since we have shown the EH Lagrangian leads to the infinite self-energy for a point charge in higher dimensions. In general, the BI-type Lagrangians do not behave as ${{\cal F}^{n + 1}}(n \ge 1)$ polynomials in the ${\cal F} \to \infty$ limit and this criterion cannot be extended to BI-type family of theories.}  

All this leads us to a definite conclusion that can be considered as a straightforward, simple criterion for the self-energy problem in nonlinear $U(1)$ gauge theories:\\
``\textit{In arbitrary dimensions, if the electrostatic field at distances very close  to the point charge grows slower than $\frac{1}{r}$, it necessarily leads to finite self-energy.}''

\begin{figure}[!htbp]
	\begin{center}
		\epsfxsize=10 cm 
		\includegraphics[width=7 cm]{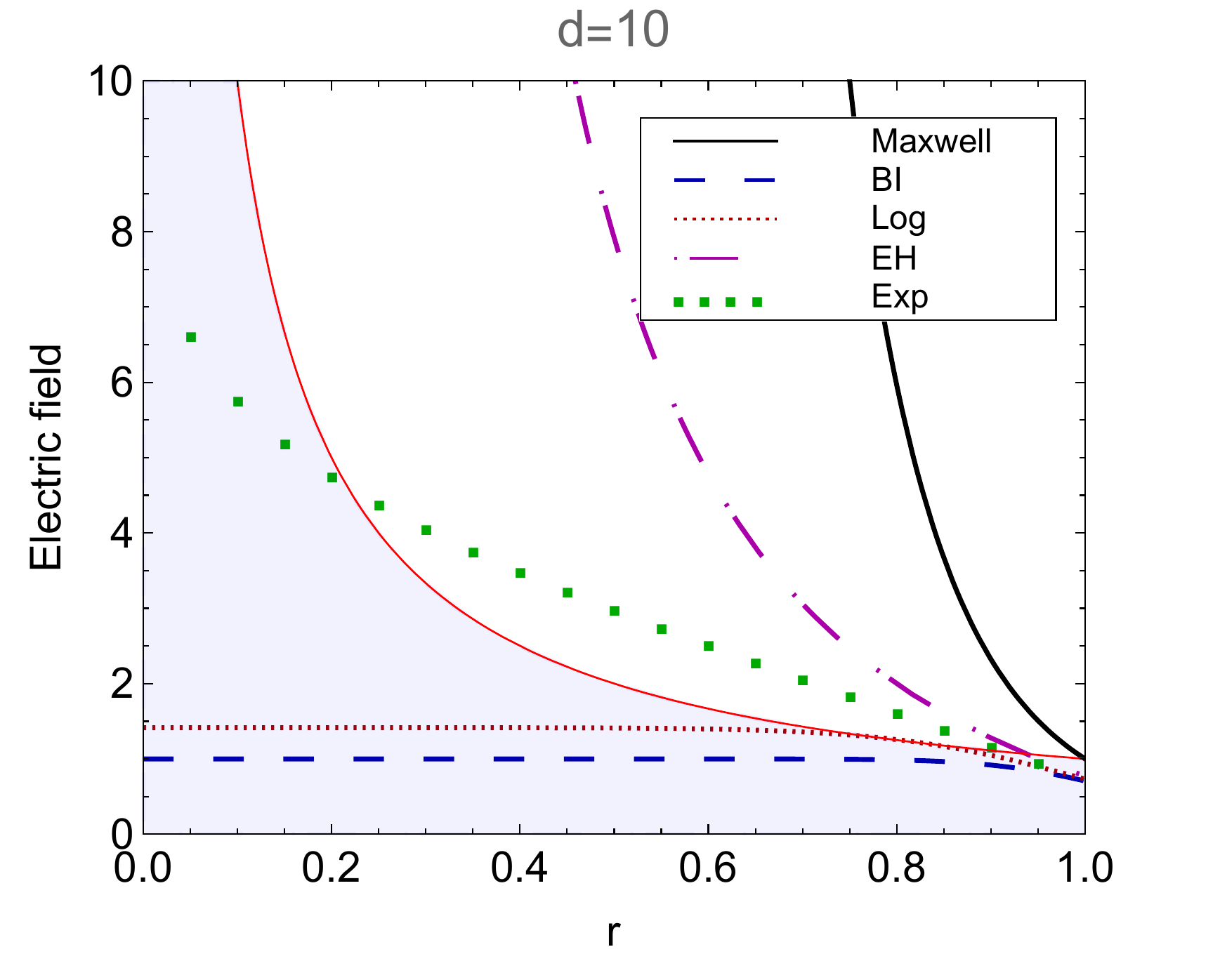}
		\hskip 0.1 cm
		\includegraphics[width=7 cm]{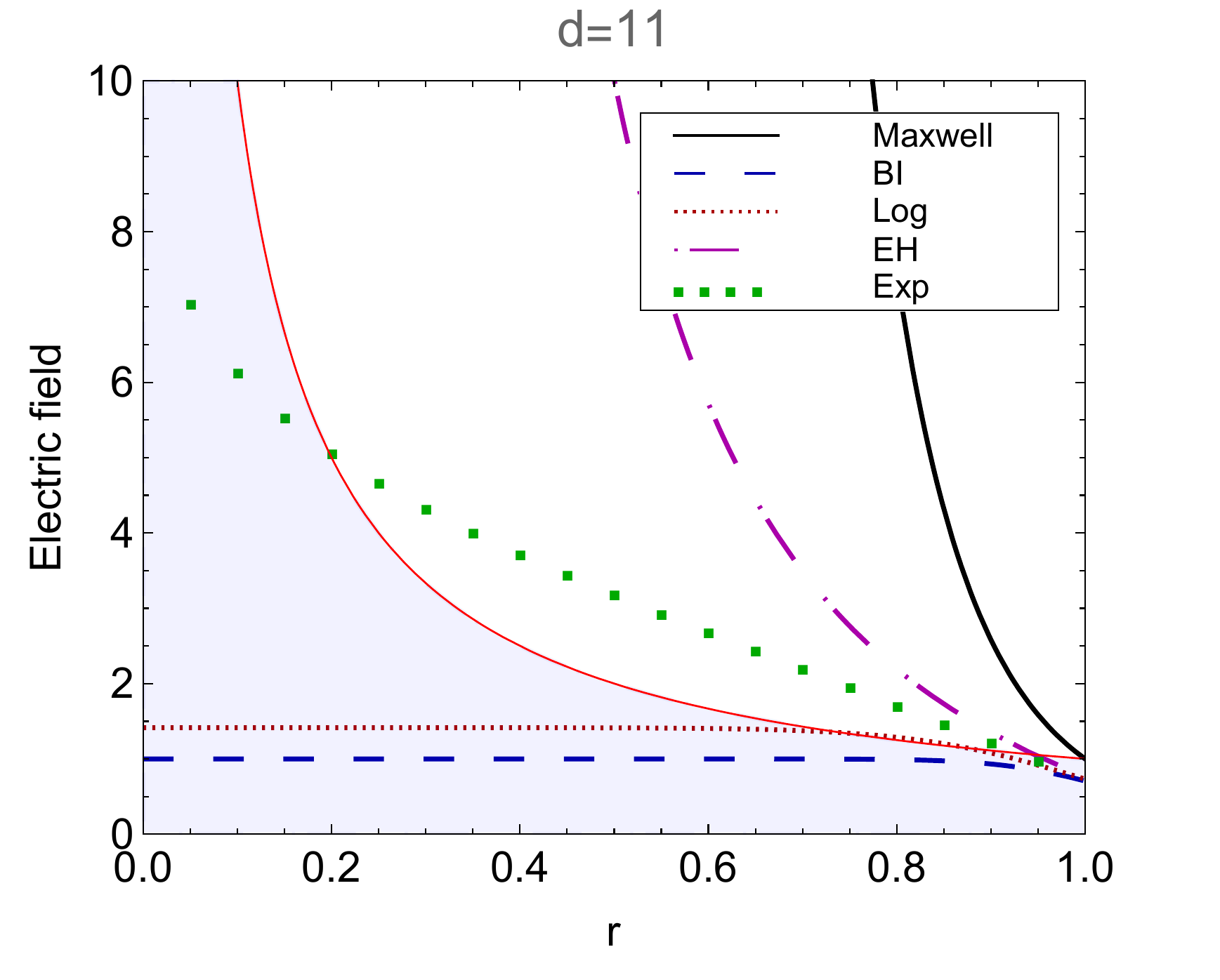}
		\caption{The behavior of the electric field in the Born-Infeld (BI), logarithmic, exponential and Euler-Heisenberg (EH) theories in $d=10$ and $11$ dimensions. The solid red lines represent the troublesome factor $\frac{1}{r}$.}
		\label{electric field_d10_11_all}
	\end{center}
\end{figure}

When we go to higher dimensions, we observe that the electric field near the origin in EH theory tends to exceed the $\frac{1}{r}$ behavior and grows rapidly. This can also be understood by taking the large-$d$ limit of the electric field in EH theory (around origin), as should be evident from figures \ref{electric field_d4_7_all} and \ref{electric field_d10_11_all}. In these figures, the solid red lines represent that troublesome factor, i.e., $\frac{1}{r}$. BI-type theories for which Taylor expansions of the electric fields at the origin grow slower than $\frac{1}{r}$, are safe and always result in finite self-energy. But, the point is that the number of BI-type theories that result in finite self-energy decreases as the number of dimensions of spacetime increases.

Here the question may arise as to whether it is legitimate to explore the self-energy problem of point charges in the weak-field truncation of the EH theory, since the strong-field behaviour of the electric field should be considered. As mentioned, the divergent behavior is rooted in the Coulomb's inverse-square law, which is the characteristic of Maxwell theory. In the weak-field limit, BI-type models and EH theory are getting closer and closer to the outcomes of the Maxwell's electrodynamics. So, if the theory is safe in the weak-field limit (i.e., results in a finite self-energy), in principle it will be safe in the strong-field limit. So, the finiteness of the self-energy for the weak-field limit of EH theory ensures that the strong-field limit of this theory results in a finite self-energy for point charges as well. But, for an accurate calculation, one needs the effective action of EH theory in the strong-(purely electric) field limit and this is still an open and important question. This issue and its obstacles have been further discussed in detail in ref. \cite{Shabad2015}.

\section{Vacuum polarization effects and comparison with QED} \label{sec:vacuum polarization}

In order to study the vacuum polarization in any classical field theory of electrodynamics, it is necessary to examine the electric permittivity and the magnetic permeability of the vacuum. In vacuum, if there is no polarization, we must actually reach the usual constants of the electric permittivity and the magnetic permeability for the vacuum that Maxwell's classical electrodynamics predicts, i.e., the well-known constants $\varepsilon_0$ and $\mu _0$ \cite{JacksonBook}. If by emitting electromagnetic fields in vacuum these quantities exceed the Maxwell limit and have an isotropic or anisotropic nature, then it will be evidence of vacuum polarization. In order to investigate vacuum polarization effects in classical electrodynamics or QED, the electric displacement field ($\bf{D}$) and the magnetic field strength ($\bf{H}$) have to be expressed in terms of electric permittivity and the inverse magnetic permeability tensors of the vacuum, given by
\begin{equation} \label{tensorial forms}
{D_i} = \sum\limits_j {{\varepsilon _{ij}}{E_j}}, \quad {H_i} = \sum\limits_j {{{\left( {{\mu ^{ - 1}}} \right)}_{ij}}{B_j}}.
\end{equation}
In what follows, we first study the vacuum polarization effects within the framework of the BI family of theories and then compare the relevant results with those of QED's.

\subsection{The polarization and magnetization of vacuum in BI-type theories}

%%If we are interested in anisotropic vacuum polarization, which may occur due to differences in field strength in different directions), ... %%

Using the definition (\ref{tensorial forms}) together with the general form of the displacement field relation (\ref{D field}), we arrive at the following relation for the electric permittivity tensor
\begin{equation}
{\varepsilon _{ij}} = \left( {{\delta _{ij}} + \frac{{2\gamma }}{{{\beta ^2}}}{B_i}{B_j}} \right)\left( { - \frac{{\partial {\cal L}}}{{\partial {\cal S}}}} \right).
\end{equation}
As seen in the previous section, the nonlinear parameter $\beta$ necessarily has a large value, which is typically comparable to the Schwinger limit. In addition, empirical observations impose the same conclusion. Expanding ${\varepsilon _{ij}} $ for large enough values of nonlinear parameter ($\beta \gg 1$) leads to

\begin{equation}
{\varepsilon _{ij}} \simeq \left( {{\delta _{ij}} + \frac{{2\gamma }}{{{\beta ^2}}}{B_i}{B_j}} \right) \times \left( {1 - {c_1}\frac{{\cal S}}{{{\beta ^2}}} + {c_2}\frac{{{{\cal S}^2}}}{{{\beta ^4}}} + ...} \right),
\end{equation}
where the coefficient $c_1$ is always the same for all BI-type theories and equals to one ($c_1 = 1$). Now, by explicitly writing the Maxwell invariant $\cal S$ in terms of $\bf{E}$ and $\bf{B}$, and then by collecting terms, one obtains
\begin{equation} \label{electric permittivity}
{\varepsilon _{ij}} \simeq {\delta _{ij}} + \frac{1}{{{\beta ^2}}}\left( {\frac{{{c_1}}}{2}\left( {{{\bf{E}}^2} - {{\bf{B}}^2}} \right){\delta _{ij}} + 2\gamma {B_i}{B_j}} \right) + O\left( {{\beta ^{ - 4}}} \right).
\end{equation}

In order to find the magnetic permeability tensor, we repeat the same exercise for the magnetic field strength (\ref{H field}) and the definition of magnetic field in term of the magnetic permeability tensor, yielding
\begin{equation}
{\left( {{\mu ^{ - 1}}} \right)_{ij}} = \left( {{\delta _{ij}} - \frac{{2\gamma }}{{{\beta ^2}}}{E_i}{E_j}} \right)\left( { - \frac{{\partial {\cal L}}}{{\partial {\cal S}}}} \right).
\end{equation}
The above relation for large enough values of $\beta$ can be expanded, leading to
\begin{equation}
{\left( {{\mu ^{ - 1}}} \right)_{ij}} \simeq \left( {{\delta _{ij}} - \frac{{2\gamma }}{{{\beta ^2}}}{E_i}{E_j}} \right) \times \left( {1 - {c_1}\frac{{\cal S}}{{{\beta ^2}}} + {c_2}\frac{{{{\cal S}^2}}}{{{\beta ^4}}} + ...} \right).
\end{equation}
Inverting this relation, the magnetic permeability tensor is found as
\begin{equation} \label{magnetic permeability}
{{\mu _{ij}} \simeq {\delta _{ij}} + \frac{1}{{{\beta ^2}}}\left( {\frac{{{c_1}}}{2}\left( {{{\bf{B}}^2} - {{\bf{E}}^2}} \right){\delta _{ij}} + 2\gamma {E_i}{E_j}} \right) + O\left( {{\beta ^{ - 4}}} \right)}.
\end{equation}

Now, let us explicitly compute the polarization and magnetization of the classical vacuum. Using the definition ${{\bf{P}} = {\bf{D}} - {\varepsilon _0}{\bf{E}}}$, we have
\begin{equation}
{\bf{P}} = (\varepsilon  - {\varepsilon _0}){\bf{E}}\, \to \,{P_i} = ({\varepsilon _{ij}} - {\delta _{ij}}){E_j}.
\end{equation}
From the above relation, one can immediately derive the induced vacuum polarization density ($\bf{P}$) as
\begin{equation} \label{P field}
{\bf{P}} = \frac{1}{{{\beta ^2}}}\left( {\frac{{{c_1}}}{2}\left( {{{\bf{E}}^2} - {{\bf{B}}^2}} \right){\bf{E}} + 2\gamma ({\bf{E}}.{\bf{B}}){\bf{B}}} \right) + O\left( {{\beta ^{ - 4}}} \right).
\end{equation}
In the same way, using the definition of magnetization vector field as ${{\bf{M}} = \mu _0^{ - 1}{\bf{B}} - {\bf{H}}}$, we have
\begin{equation}
{\bf{M}} = \left( {\mu _0^{ - 1} - {\mu ^{ - 1}}} \right){\bf{B}}\, \to \,\,{M_i} = \left( {{\delta _{ij}} - {{\left( {{\mu ^{ - 1}}} \right)}_{ij}}} \right){B_j}.
\end{equation}
Therefore, the magnetization vector field is given by
\begin{equation} \label{M field}
{\bf{M}} =  - \frac{1}{{{\beta ^2}}}\left( {\frac{{{c_1}}}{2}\left( {{{\bf{B}}^2} - {{\bf{E}}^2}} \right){\bf{B}} + 2\gamma ({\bf{B}}.{\bf{E}}){\bf{E}}} \right) + O\left( {{\beta ^{ - 4}}} \right).
\end{equation}

So far, we have shown the electric permittivity and the magnetic permeability of the vacuum should be considered as tensors which indicates the classical vacuum has an anisotropic and nonlinear nature. The inclusion of the Maxwell invariant ${\cal G}$ ($= \frac{1}{4}{F_{\mu \nu }}{}^*{F^{\mu \nu }}$) in the Lagrangian density is essential to obtain the latter result. If we want to have an isotropic vacuum polarization, the Maxwell invariant $\cal G$ must vanish (simply by setting $\gamma = 0$). In the case of isotropic vacuum, it turns out that the displacement field and the magnetic field strength may be written as
\begin{equation}
{\bf{D}} = \frac{{\partial {\cal L}}}{{\partial {\bf{E}}}} = \left( { - \frac{{\partial {\cal L}}}{{\partial S}}} \right){\bf{E}},
\end{equation}
and
\begin{equation}
{\bf{H}} =  - \frac{{\partial {\cal L}}}{{\partial {\bf{B}}}} = \left( { - \frac{{\partial {\cal L}}}{{\partial S}}} \right){\bf{B}}.
\end{equation}
Using the standard definitions of ${\bf{D}} = \varepsilon {\bf{E}} = {\varepsilon _0}{\varepsilon _{\rm{r}}}{\bf{E}}$ and ${\bf{H}} = {\mu ^{ - 1}}{\bf{B}} = {\left( {{\mu _0}{\mu _{\rm{r}}}} \right)^{ - 1}}{\bf{B}}$ together with eq. (\ref{asymp-BI-EH}), the quantities $\varepsilon$ and $\mu$ are found to be scalars and are related through the following relation
\begin{eqnarray}
\frac{\varepsilon }{{{\varepsilon _0}}} = \frac{{{\mu _0}}}{\mu } &=&  - \frac{{\partial {\cal L}}}{{\partial {\cal F}}}\\ \nonumber
&\simeq& 1 - {c_1}\frac{{\cal F}}{{{\beta ^2}}} + {c_2}\frac{{{{\cal F}^2}}}{{{\beta ^4}}} + O\left( {{\beta ^{ - 6}}} \right).
\end{eqnarray}
%Of course, one can still use the tensorial form of electric permittivity and the magnetic permeability but, in such situations, these tensors are diagonal with equal components.
Isotropic polarization indicates that the relevant properties are independent of the direction of examination. So, the classical vacuum within the context of nonlinear electrodynamics behave much like optically isotropic materials if ${\cal G}=0$. The polarization and the magnetization are simply obtained by setting $\gamma = 0$ in eq. (\ref{P field}) and eq. (\ref{M field}), respectively.

\subsection{Comparison with the results of QED}

The obtained results in the previous section are very interesting since they clearly exhibit the polarization effects of the vacuum in a classical way. At this stage, a connection can be made with the vacuum polarization effects in QED. So, let us compare these results with those of QED.

In QED, the vacuum polarization effects are due to the one-loop contributions of fermion–antifermion pairs in photon propagator \cite{KlauberQFT,SchwartzQFT,MandlShaw}. One can think of it in this way: the background electromagnetic field may produce virtual particle-antiparticle pairs with the opposite electric charges and, if there is a charge in empty space, the electromagnetic field of the physical charge disrupts the particle-antiparticle pair distribution. As a result, this new distribution of virtual particle-antiparticle pairs alters the initial electromagnetic field. As a consequence, the vacuum also shows polarization effects and therefore the corresponding tensors of the electric permittivity and the magnetic permeability  will not be some constants anymore. For these tensorial quantities, the QED results in \cite{LifshitzQED,BerestetskiiQED}
\begin{equation} \label{QED permittivity}
{\varepsilon _{ij}} = {\delta _{ij}} + \frac{{{e^4}\hbar }}{{45\pi m_e^4{c^7}}}\left[ {2\left( {{{\bf{E}}^2} - {{\bf{B}}^2}{c^2}} \right){\delta _{ij}} + 7{c^2}{B_i}{B_j}} \right] + {\rm{higher}}\,\,{\rm{orders}}\,{\rm{,}}
\end{equation}
and
\begin{equation} \label{QED permeability}
{\mu _{ij}} = {\delta _{ij}} + \frac{{{e^4}\hbar }}{{45\pi m_e^4{c^7}}}\left[ {2\left( {{{\bf{B}}^2}{c^2} - {{\bf{E}}^2}} \right){\delta _{ij}} + 7{c^2}{E_i}{E_j}} \right] + {\rm{higher}}\,\,{\rm{orders}}\,{\rm{.}}
\end{equation}
Comparing these with the results of BI-type nonlinear theories for the electric permittivity and the magnetic permeability tensors in vacuum, i.e., eq. (\ref{electric permittivity}) and eq. (\ref{magnetic permeability}), it is clearly seen that the vacuum polarization effects have the same mathematical form. In nonlinear electrodynamics, usually, $\gamma$ is set as $\gamma = \frac{1}{2}$, but this choice is very limited and restricts the theory significantly. What is somewhat surprising is that since $c_1=1$ (this is always possible) by setting $\gamma = \frac{7}{8}$ an exact one-to-one correspondence with QED and the effective EH theory is obtained up to the leading order of corrections. Consequently, the coincidence with the QED results of the vacuum polarization effects is more direct and precise. Now, it should be obvious that why the inclusion of the fundamental invariance $\cal G$ in Lagrangian density is necessary in order to get the consistent result with those of QED. Equating (\ref{electric permittivity}) to (\ref{QED permittivity}), the nonlinear parameter $\beta$ for any BI-type theory is exactly (not approximately) identified with the basic constants of QED as

\begin{equation} \label{beta}
\frac{1}{{4{\beta ^2}}} = \frac{{{e^4}\hbar }}{{45\pi m_e^4{c^7}}}\,\,\, \Rightarrow \,\,\beta  = \frac{{3m_e^2{c^3}}}{{2{e^2}}}\sqrt {\frac{{5\pi c}}{\hbar }},
\end{equation}
with the numerical value in SI units as
\begin{equation} \label{BI-type beta}
\beta_{\rm{BI-type}}=9.20908232321942 \times 10^{19} \,{\rm{V/m}}.
\end{equation}
Again, this is slightly greater than the Schwinger scale (\ref{Schwinger limit}), as it should be.\footnote{Interestingly, this value is very close to the one that obtained for the logarithmic model in ref. \cite{Gaete2014a} (repeated in eq. (\ref{theoretical beta}) in the present work).} Since this computation is valid in the weak-field limit, it does not matter which BI-type theory we use. As it is evident, the value obtained here for the nonlinearity parameter is in agreement with what we have found from the electron's self-energy in eqs. (\ref{exp beta}) and (\ref{theoretical beta}).

\section{Dual symmetry} \label{sec:duality}

In the context of classical electrodynamics, it has been shown that the sourceless Maxwell equations satisfy a simple constraint as \cite{Schrodinger1935}
\begin{equation} \label{duality constraint}
{\bf{D}}{\rm{.}}{\bf{H}} = {\bf{E}}{\rm{.}}{\bf{B}}.
\end{equation}
When the condition (\ref{duality constraint}) is satisfied, the sourceless Maxwell's equations remain invariant under the $SO(2)$ electromagnetic duality group \cite{Gibbons1995,Gibbons1996}. This is the dual symmetry (also known as the electric-magnetic duality) and any dual symmetric $U(1)$ gauge theory must satisfy this constraint. 

Now let us look at this problem in BI-family of nonlinear $U(1)$ gauge theories. We first assume that both the Maxwell invariants $\cal F$ and $\cal G$ appear in the invariant combination as ${\cal S} = {\cal F} - \frac{{\gamma {{\cal G}^2}}}{{{\beta ^2}}}$, which was already supported by the effective EH theory due to the vacuum polarization effects (see section \ref{sec:vacuum polarization}). Using the dot product of the D-field (\ref{D field}) with the H-field (\ref{H field}), one can straightforwardly show that 
\begin{equation} 
{\bf{D}}{\rm{.}}{\bf{H}} = ({\bf{E}}{\rm{.}}{\bf{B}})\left( {1 + \frac{{4\gamma }}{{{\beta ^2}}}{\cal S}} \right){\left( {\frac{{\partial {\cal L}}}{{\partial {\cal S}}}} \right)^2}.
\end{equation}
From this, it is inferred that the necessary and sufficient condition for the BI-type theories to have this symmetry is
\begin{equation} \label{DS condition}
\left( {1 + \frac{{4\gamma }}{{{\beta ^2}}}{\cal S}} \right){\left( {\frac{{\partial {\cal L}}}{{\partial {\cal S}}}} \right)^2} = 1.
\end{equation}
Solving the latter condition, we arrive at
\begin{equation} \label{general L}
\left( {\frac{{\partial {\cal L}}}{{\partial {\cal S}}}} \right) =\pm \frac{1}{{\sqrt {1 + \frac{{4\gamma }}{{{\beta ^2}}}{\cal S}} }}\,\,\,\, \Rightarrow \,\,\,\,{\cal L} =  \pm \frac{{{\beta ^2}}}{{2\gamma }}\sqrt {1 + \frac{{4\gamma }}{{{\beta ^2}}}{\cal S}}  + C,
\end{equation}
where $C$ is an integration constant. The constant $C$ is fixed by demanding that in the weak-field limit ($\beta \to \infty$) the Lagrangian must reduce to the physical case of Maxwell's theory. The weak-field expansion of the obtained Lagrangian (\ref{general L}) takes the following form
\begin{equation}
{\cal L} =  \pm \frac{{{\beta ^2}}}{{2\gamma }} \pm {\cal S} + C \mp \frac{{\gamma {{\cal S}^2}}}{{{\beta ^2}}} + O\left( {{\beta ^{ - 4}}} \right).
\end{equation}
When comparing with the Lagrangian density of classical electrodynamics, this clearly shows the obtained Lagrangian (\ref{general L}) with the positive sign is never acceptable. If the minus sign is chosen, the Lagrangian reduces to the Maxwell's case, which leads to $C = \frac{{{\beta ^2}}}{{2\gamma }}$. In conclusion, the most general version of dual symmetric Lagrangian is obtained as 
\begin{equation} \label{DBI}
{{\cal L}_{{\rm{EM-duality}}}} = \frac{{{\beta ^2}}}{{2\gamma }}\left( {1 - \sqrt {1 + \frac{{4\gamma }}{{{\beta ^2}}}\left( {{\cal F} - \frac{{\gamma {\cal G}}}{{{\beta ^2}}}} \right)} } \right)
\end{equation}
In the original BI theory, the authors introduced their theory’s Lagrangian as \cite{Born-Infeld1934}
\begin{equation}
{{\cal L}_{{\rm{BI}}}} = {\beta ^2}\left( {1 - \sqrt {1 + \frac{{{{\bf{B}}^2} - {{\bf{E}}^2}}}{{{\beta ^2}}} - \frac{{{\bf{E}}.{\bf{B}}}}{{{\beta ^4}}}} } \right),
\end{equation}
which is equivalent to the case with $\gamma  = \frac{1}{2}$ in eq. (\ref{DBI}). A simple redefinition of the nonlinear parameter as $\frac{{{\beta ^2}}}{{2\gamma }} \to {\beta ^2}$ also gives the same result. This simple proof indicate that the BI theory, eq. (\ref{DBI}), is the only dual symmetric nonlinear $U(1)$ gauge theory with the asymptotic behavior (\ref{asymp-BI-EH}). So, it is then not so surprising that why the other types of BI-type nonlinear $U(1)$ gauge theories are not dual symmetric. Interestingly, the electric-magnetic duality was already found to be absent in different models of BI-type theories by case by case studies (e.g., see refs. \cite{Kruglov2016,DoubleLogaritmicNED2021,Kruglov2017} and references therein), giving the dual symmetric BI theory (\ref{DBI}) a special significance. 

The important conclusion is that, in order to retain the electric-magnetic duality invariance of the Lagrangian, the Maxwell invariants $\cal F$ and $\cal G$ must always appears in a specific combination, which is equivalent to the case with $\gamma  = \frac{1}{2}$ in eq. (\ref{DBI}) (Again, this can simply be inferred by the redefinition $\frac{{{\beta ^2}}}{{2\gamma }} \to {\beta ^2}$.) This choice would ruin the one-to-one correspondence of vacuum polarization effects in BI-type theories with QED and the effective EH theory (up to the leading order of corrections) that we have already obtained in the previous section \ref{sec:vacuum polarization}. In section \ref{sec:vacuum polarization}, we have proved that the coefficient $\gamma$ in front of the Maxwell invariant $\cal G$ (not an overall $\gamma$ like that in eq. (\ref{DBI})) has to be set as $\gamma = \frac{7}{8}$. Furthermore, it should be noted that the weak-field limit of BI-type theories does not satisfy the essential criterion (\ref{DS condition}). It seems the dual symmetry is always broken in the weak-field expansion of nonlinear $U(1)$ gauge theories due to quantum mechanical corrections.

It should be emphasized that, theoretically, it is quite possible to preserve dual symmetry in more general $U(1)$ gauge theories. Here, we restricted ourself to the BI family of $U(1)$ gauge theories with the EH weak-field limit (\ref{asymp-BI-EH}). Interestingly, ModMax theory, which has been recently introduced in refs. \cite{ModMax2020PRD,ModMax2021JHEP}, preserves both dual symmetry and conformal invariance. Evidently, this theory does not fit into our classification for BI-type theories and, consequently, it does not incorporate the exact forms of QED's vacuum polarization effects (\ref{QED permittivity}) and (\ref{QED permeability}), which are exactly explained by the weak-field limit of EH and BI-type theories. Naturally, it is possible to generalize BI theory with some new parameters in order to build new dual symmetric models. In ref. \cite{Hatsuda1999}, the authors have presented some examples of Lagrangians which are one-parameter families of generalized Born-Infeld theories. The Lagrangians introduced in eqs. (4.15), (4.23), and (4.27) of ref. \cite{Hatsuda1999} all reduce to BI Lagrangian for $a \to 0$. However, for the finite value of $a$ they have different weak-field expansions.

\section{Summary and discussion} \label{sec:conclusion}

Nonlinear classical field theories of electrodynamics have been devised to resolve some limitations of
Maxwell's electrodynamics (such as the infinite self-energy of point charges, the absence of light by light scattering etc) without incorporating the principles of quantum physics. These theories
generally are referred to as nonlinear electrodynamics. It is seen that all the BI \cite{Born-Infeld1934}, EH \cite{HeisenbergEuler1936}, logarithmic \cite{Soleng} and exponential \cite{Hendi2012} theories of nonlinear electrodynamics at the weak-field limit have mathematically the same form as eq. (\ref{asymp-BI-EH}). Motivated by this fact, we have studied a family of nonlinear $U(1)$ gauge theories which exactly matches the weak-field limit of both BI and EH theories. This condition guarantees that the outcomes of Maxwell's classical electrodynamics are recovered in the weak-field limit. This family of theories are often referred to as BI-type models of nonlinear electrodynamics. In the weak-field limit, these theories schematically have the same form as eq. (\ref{asymp-BI-EH}). But the exact functional form of the Lagrangian density is important in strong-field coupling limit, where the expansion (\ref{asymp-BI-EH}) is not allowed. The analyses presented here have the significant advantage of being a simple and tractable treatment for studying BI-type family of nonlinear $U(1)$ gauge theories.

We have first concentrated on the self-energy problem in BI-type family of theories. For this purpose, we have started with the unsolved case of exponential nonlinear electrodynamics \cite{Hendi2012,Hendi2013}. Regarding the exponential model, it has explicitly proved that although the electric field at the location of the elementary point charges is infinite, but the total electrostatic field energy is finite. This is a remarkable finding since the same happens for the weak-field limit of EH theory in which the field energy of a point charge is finite as well \cite{Shabad2015}. The proof of the finiteness of the self-energy in BI, EH, logarithmic and exponential nonlinear $U(1)$ gauge theories has also been extended to higher dimensional spacetimes. (This is important in the subjects of gauge/gravity duality, black hole physics and brane cosmology where higher dimensions play a key role.) Interestingly, it was proved that the EH theory cannot resolve the self-energy
issue in higher dimensions. Since the effective EH theory is the weak-field limit of BI-type theories, we achieve the following results:

\begin{itemize}
	\item It is conjectured that \textit{all BI-type nonlinear $U(1)$ gauge theories lead to a finite self-energy for the elementary point charges in 4-dimensions}. While they classically preserve the notion of \textit{point charge} well.
	
	\item In 5-dimensions, all the BI family of theories result in finite self-energy except for the weak-field limit of these family of theories, i.e., the weak-field limit of EH theory.
	
	\item As the number of dimensions of spacetime increases, the number of BI-type theories that result in finite self-energy decreases.
	
	\item In arbitrary dimensions, there exist some BI-type theories in which self-energy is always finite. BI, logarithmic, and exponential models are three explicit examples.
	
	\item For any nonlinear $U(1)$ gauge theory, if the electrostatic field at very close distances to the point charge grows at a rate less than $\frac{1}{r}$, it necessarily leads to a finite self-energy. Remarkably, this criterion is valid in arbitrary dimensions.
	
\end{itemize}

By construction, the nonlinear Lagrangians based on the basic radical or logarithmic functions lead to a maximum on the electric field. So, they trivially results in finite self-energy.\footnote{The explicit examples are BI \cite{Born-Infeld1934}, logarithmic \cite{Soleng} and double-logarithmic \cite{DoubleLogaritmicNED2021} models of nonlinear electrodynamics.} In these kinds of BI-type theories, $\beta$ is proportional to the maximum possible electric field. (The value of $\beta$ is determined by computing the self-energy of the point charge, e.g., the electron. Generally, the order of magnitude is obtained as $\beta_{\rm{BI-type}} \approx 10^{19}-10^{20} \rm{V}/\rm{m}$.) As seen in section \ref{sec:self-energy}, leaving aside this assumption and considering other basic functions for Lagrangian (e.g., exponential or quadratic forms), one can still obtain a finite value for the self-energy of point charges with $\beta \approx 10^{19} \rm{V}/\rm{m}$. In these kinds of BI-type theories, the fields can exceed the critical value $\beta$. What is almost certain is that there is a general agreement that the maximum possible electric field must be greater than that proposed by BI theory\footnote{Different theoretical and experimental pieces of evidence indicate that the value of the nonlinear parameter ($\beta$) should be (perhaps much) larger than the values predicted by BI theory, e.g.,
		\begin{itemize}
			\item  in ref. \cite{Rafelski1973}, the authors have found the lower limit on the BI mass scale $M=\sqrt{\beta}>100$ MeV that is equivalent to $\beta \ge 1.7 \times 10^{22} \rm{V}/\rm{m}$   from muonic transitions in lead. This value is two orders of magnitude larger than the original BI parameter in Eq. (\ref{theoretical beta}). 
			
			\item in ref. \cite{Akmansoy2018}, the hydrogen’s ionization energy
			was used to set a lower limit on the nonlinearity parameter $\beta$ as $\beta \ge 10^{21} \rm{V}/\rm{m}$ for the BI, logarithmic and exponential models. 
			
			\item in ref. \cite{Ellis2017}, the authors have found the lower limit on the BI mass scale $M=\sqrt{\beta}>100$ GeV that is equivalent to $\beta \ge 4.3 \times 10^{27} \rm{V}/\rm{m}$ from the measurement of light-by-light scattering in LHC pb-pb collision \cite{ATLAS2017}.
		\end{itemize} 
		Nevertheless, in ref. \cite{Davila2014}, the lower bound on $\beta $  as $\beta  > 2 \times {10^{19}}{\rm{V/m}}$  was found from theoretical study of a purely photonic process, namely photon splitting in a magnetic field in BI theory and then comparing with the measurements of photon splitting in neutron stars \cite{Enoto}. This bound is approximately around the original BI nonlinear parameter in Eq. (\ref{theoretical beta}) and three orders of magnitude smaller than what is found in \cite{Rafelski1973}. 
		} , e.g., the electromagnetic fields as large as $10^{25} \rm{V}/\rm{m}$ has been announced by ATLAS \cite{ATLAS2017}.

Next, we focused on the effects of classical vacuum polarization and magnetization in BI-type theories. It has been shown that there is a systematic way to reveal the polarization of the classical vacuum in all BI-type theories. It has been shown that the inclusion of the Maxwell invariant ${\cal G}$($={F_{\mu \nu }}{}^*{F^{\mu \nu }}/4$) in the Lagrangian densities of BI-type theories (using the replacement (\ref{S invariant})) leads to vacuum polarization effects, structurally the same as QED results. We have shown that, generally, all the BI-type models of nonlinear electrodynamics predict the vacuum polarization effects in a classical way, exactly in one-to-one correspondence with QED and the effective EH theory up to the leading order of corrections. In conclusion, the effect of QED radiative corrections can naturally be simulated by classical theories of BI-type nonlinear electrodynamics. 

In addition to the proof in ref. \cite{Gibbons1995}, we have presented a new, simpler proof about the absence of dual symmetry (electric-magnetic duality) in BI-type nonlinear $U(1)$ gauge theories, in which the original BI theory is quite exceptional and preserves this symmetry like Maxwell's theory. Since the dual symmetry is preserved under the condition (\ref{DS condition}), therefore, in order to have the dual symmetry, the Maxwell invariant $\cal G$ cannot vary independent of $\cal F$ and they always must appear in the invariant combination of $\frac{{4\gamma }}{{{\beta ^2}}}( {{\cal F} - \frac{{\gamma {\cal G}}}{{{\beta ^2}}}})$. This implies a very important result: it is not possible to simultaneously have both the properties of vacuum polarization  (exactly in one-to-one correspondence with QED and EH theory up to the leading order of corrections) and dual symmetry in a classical $U(1)$ gauge theory of nonlinear electrodynamics.

\begin{table}[]
	\caption{Summary of different models of nonlinear electrodynamics with their properties.}
	\label{tab:summary}
	\begin{tabular}{|c|c|c|c|c|}
		\hline
		\textit{Theory}                                                                 & \textit{\begin{tabular}[c]{@{}c@{}}Maximum\\ electric field\end{tabular}} & \textit{\begin{tabular}[c]{@{}c@{}}Self-energy\\ of point charges\\
		in 4-dimensions\end{tabular}} & \textit{\begin{tabular}[c]{@{}c@{}}Vacuum\\ birefringence\end{tabular}} & \textit{\begin{tabular}[c]{@{}c@{}}Vacuum\\ polarization\end{tabular}} \\ \hline
		\begin{tabular}[c]{@{}c@{}}Born-Infeld\\ electrodynamics\end{tabular}           & yes                                                                       & \begin{tabular}[c]{@{}c@{}}finite \cite{Born-Infeld1934}\\ ${\cal E}_{\text{BI}}=1.236 \sqrt{\beta Q^3}$\end{tabular}                             & no                                                                      & yes                                                                    \\ \hline
		\begin{tabular}[c]{@{}c@{}}Euler-Heisenberg\\ electrodynamics\end{tabular}      & no                                                                        & \begin{tabular}[c]{@{}c@{}}finite \cite{Shabad2015}\\ ${\cal E}_{\text{EH}}=2.940 \sqrt{\beta Q^3}$\end{tabular}                             & yes \cite{Kruglov2007}                                                                    & yes                                                                    \\ \hline
		\begin{tabular}[c]{@{}c@{}}Logarithmic nonlinear\\ electrodynamics\end{tabular} & yes                                                                       & \begin{tabular}[c]{@{}c@{}}finite \cite{Gaete2014a}\\ ${\cal E}_{\text{log}}=1.386 \sqrt{\beta Q^3}$\end{tabular}                            & yes   \cite{Gaete2014a}                                                                  & yes                                                                    \\ \hline
		\begin{tabular}[c]{@{}c@{}}Exponential nonlinear\\ electrodynamics\end{tabular} & no                                                                        & \begin{tabular}[c]{@{}c@{}}finite\\ ${\cal E}_{\text{exp}}=1.709 \sqrt{\beta Q^3}$\end{tabular}                            & yes   \cite{Gaete2014b}                                                                  & yes                                                                    \\ \hline
	\end{tabular}
\end{table}

In conclusion, the problems of electron's self-energy, vacuum polarization effects and light-by-light scattering can be addressed within the BI-type nonlinear $U(1)$ gauge theories in a satisfactory way. For convenience, we have summarized some parts of the results in table \ref{tab:summary}. From this table, it is inferred that the results of the exponential model are more similar to the results of EH theory, while they are different in some aspects from BI and logarithmic models of nonlinear electrodynamics. Furthermore, both the self-energy problem and vacuum polarization effects can appropriately be explained by every model of BI family of nonlinear $U(1)$ gauge theories.

% Extra notes:
% We argue that such models suffer from arbitrariness; 

\acknowledgments

S.Z. appreciates the support of University of Sistan and Baluchestan's research council. A.D. wish to thank the support of Iran Science Elites Federation (ISEF).\\

\paragraph{Data Availability Statement} Data sharing is not applicable, as no data sets were generated or analyzed in this
research. (This is a theoretical work and no experimental data were used.)


\begin{thebibliography}{99}


\bibitem{Born-Infeld1934}M. Born and L. Infeld, \textit{Foundations of the new field theory}, Proc. Roy. Soc. Lond. A \textbf{144} (1934) 425

\bibitem{Einstein1905}A. Einstein, \textit{On the electrodynamics of moving bodies}, Annalen der Physik \textbf{17} (891) 50

\bibitem{Heisenberg1938} W. Heisenberg, \textit{Über die in der Theorie der Elementarteilchen auftretende universelle Länge}, Ann. Physik \textbf{32} (1938) 20

\bibitem{MoayediSetare2012}S.K. Moayedi, M.R. Setare, and H. Moayeri, \textit{Formulation of an electrostatic field with a charge density in the presence of a minimal length based on the Kempf algebra}, EPL (Europhysics Letters) \textbf{98} (2012) 50001


\bibitem{Kempf1997}A. Kempf and G. Mangano, \textit{Minimal length uncertainty relation and ultraviolet regularization}, Phys. Rev. D \textbf{55} (1997) 7909


\bibitem{Hossenfelder2006}S. Hossenfelder, \textit{Interpretation of quantum field theories with a minimal length scale}, Phys. Rev. D \textbf{73} (2006) 105013


\bibitem{Feynman1964}R.P. Feynman, R.B. Leighton, and M.L. Sands, \textit{The Feynman lectures on physics}, Physics Today \textbf{17} (1964) 45

\bibitem{Soleng}H.H. Soleng, \textit{Charged black points in general relativity coupled to the logarithmic $U(1)$ gauge theory}, Phys. Rev. D \textbf{52} (1995) 6178

\bibitem{Gaete2014a}P. Gaete and J. Helayl-Neto, \textit{Finite field-energy and interparticle potential in logarithmic electrodynamics}, Eur. Phys. J. C \textbf{74} (2014) 2816

\bibitem{Hendi2012}S.H. Hendi, \textit{Asymptotic charged BTZ black hole solutions}, J. High Energy Phys. \textbf{03} (2012) 065

\bibitem{Hendi2013}S.H. Hendi, \textit{Asymptotic Reissner–Nordström black holes}, Annals of Physics \textbf{333} (2013) 282


\bibitem{Kruglov2016}S.I. Kruglov, \textit{Nonlinear electromagnetic fields as a source of universe acceleration}, Int. J. Mod. Phys. A, 31 (2016) 1650058



\bibitem{DoubleLogaritmicNED2021}I. Gullu and S.H. Mazharimousavi, \textit{Double-logarithmic nonlinear electrodynamics}, Phys. Scr. \textbf{96} (2021) 045217







\bibitem{Gaete2014b}P. Gaete and J. Helayël-Neto, \textit{Remarks on nonlinear electrodynamics}, Eur. Phys. J. C \textbf{74} (2014) 3182



% Vacuum Birefringence from Theory
\bibitem{VB2013a} K. Hattori and K. Itakura, \textit{Vacuum birefringence in strong magnetic fields: (I) Photon polarization tensor with all the Landau levels}, Ann. Phys. \textbf{330} (2013) 23

\bibitem{VB2013b}K. Hattori and K. Itakura, \textit{Vacuum birefringence in strong magnetic fields: (II) Complex refractive index from the lowest Landau level}, Ann. Phys. \textbf{334} (2013) 58

\bibitem{VB2017}V. I. Denisov, E. E. Dolgaya, and V. A. Sokolov, \textit{Nonperturbative QED vacuum birefringence},  J. High Energy Phys. \textbf{05} (2017) 105

% Vacuum Birefringence from Experiment


\bibitem{VB-exp-2013}S. Villalba-Chávez and A. Di Piazza, \textit{Axion-induced birefringence effects in laser driven nonlinear vacuum interaction}, J.  High Energy Phys. \textbf{11} (2013) 136

\bibitem{VB-exp-2014} F. Della Valle et al., \textit{First results from the new PVLAS apparatus: a new limit on vacuum magnetic birefringence}, Phys Rev. D \textbf{90} (2014) 092003

\bibitem{VB-exp-2016} F. Della Valle et al., \textit{The PVLAS experiment: measuring vacuum magnetic birefringence and dichroism with a birefringent Fabry-Perot cavity}, Eur. Phys. J. C \textbf{76} (2016) 24


\bibitem{Gibbons1995} G.W. Gibbon and D.A. Rasheed, \textit{Electric-magnetic duality rotations in non-linear electrodynamics}, Nuc. Phys. B \textbf{454} (1995) 185

\bibitem{Gibbons1996} G.W. Gibbons and D.A. Rasheed, \textit{$SL (2, {\cal R})$ invariance of non-linear electrodynamics coupled to an axion and a dilaton}, Phys. Lett. B \textbf{365} (1996) 46


\bibitem{ModMax2020PRD}I. Bandos, K. Lechner, D. Sorokin, and P.K. Townsend, \textit{Nonlinear duality-invariant conformal extension of Maxwell’s equations}, Phys. Rev. D \textbf{102} (2020) 121703


\bibitem{ModMax2021JHEP}I. Bandos, K. Lechner, D. Sorokin, and P.K. Townsend, \textit{On p-form gauge theories and their conformal limits}, J. High Energy Phys. \textbf{3} (2021) 22

\bibitem{Babaei2016} K. Babaei Velni, , and H. Babaei-Aghbolagh, \textit{On $SL (2, R)$ symmetry in nonlinear electrodynamics theories}, Nuc. Phys. B \textbf{913} (2016): 987

\bibitem{Babaei2021}H. Babaei-Aghbolagh, K. Babaei Velni, D. Mahdavian Yekta, and H. Mohammadzadeh, \textit{$T\overline {T} $-like flows in non-linear electrodynamic theories and S-duality}, J. High Energy Phys. 2021 \textbf{04} (2021) 187


\bibitem{HeisenbergEuler1936}W. Heisenberg and H. Euler, \textit{Folgerungen aus der Diracschen Theorie des Positrons}, Z. Phys. \textbf{98} (1936) 714;\\ Translated in En: \textit{Consequences of Diracs Theory of the Positron}, by W. Korolevski and H. Kleinert, arXiv:physics/0605038



\bibitem{Halpern1934} O. Halpern, \textit{Scattering processes produced by electrons in negative energy states}, Phys. Rev. \textbf{44} (1933) 885


\bibitem{Euler-Kockel1935}H. Euler and B. Kockel, \textit{The scattering of light by light in the Dirac theory}, Naturwiss. \textbf{23} (1935) 246


\bibitem{Euler1936}H. Euler, \textit{On the scattering of light by light in Dirac’s theory}, Ann. Phys. (Leipzig) \textbf{26} (1936) 398


\bibitem{KlauberQFT}R.D. Klauber, \textit{Student Friendly Quantum Field Theory: Basic Principles and Quantum Electrodynamics} (Sandtrove Press, Fairfield, IA, 2013)

\bibitem{SchwartzQFT}M.D. Schwartz, Quantum field theory and the standard model (Cambridge University Press, 2014)


\bibitem{MandlShaw}F. Mandl, Franz and G. Shaw, \textit{Quantum Field Theory}. (John Wiley \& Sons, Ltd, 2010)


\bibitem{Thomas1936}C.D. Thomas, \textit{The Scattering of Light by Light According to the Born-Infeld Theory}, Phys. Rev. \textbf{50} (1936) 1046

\bibitem{Schrodinger1941}E. Schr{\"o}dinger, \textit{Non-Linear Optics}, Proc. R. Ir. Acad. A \textbf{47} (1942) 77

\bibitem{Davila2014}J.M. Dávila, C. Schubert, and M.A. Trejo, \textit{Photonic processes in Born–Infeld theory}, Int. J. Mod. Phys. A \textbf{29} (2014) 1450174

\bibitem{Ellis2017}J. Ellis, John, N.E. Mavromatos, and T. You, \textit{Light-by-light scattering constraint on Born-Infeld theory}, Phys. Rev. Lett. \textbf{118} (2017) 261802

\bibitem{Rebhan2017}A. Rebhan and G. Turk, \textit{Polarization effects in light-by-light scattering: Euler-Heisenberg versus Born–Infeld}, Int. J. Mod. Phys. A \textbf{32} (2017) 1750053

\bibitem{ATLAS2017}ATLAS collaboration, \textit{Evidence for light-by-light scattering in heavy-ion collisions with the ATLAS detector at the LHC}, Nature Phys. \textbf{13} (2017) 852

\bibitem{StringBI1}E. S. Fradkin and A. A. Tseytlin, \textit{Non-linear electrodynamics from quantized strings}, Phys. Lett. B \textbf{163} (1985) 123

\bibitem{StringBI2} A. Abouelsaood, C. G. Callan, C. R. Nappi, and S. A. Yost, \textit{Open strings in background gauge fields}, Nuc. Phys. B \textbf{280} (1987) 599.

\bibitem{Gross1987} D.J. Gross and J.H. Sloan, \textit{The quartic effective action for the heterotic string}, Nucl. Phys. B \textbf{291} (1987) 41

\bibitem{StringBI3}A. A. Tseytlin, \textit{On non-abelian generalisation of the Born-Infeld action in string theory}, Nuc. Phys. B \textbf{501} (1997) 41.

\bibitem{StringBI4} B. Zwiebach, \textit{A first course in string theory} (Cambridge university press, 2004).


\bibitem{Beato1999}E. Ayon-Beato and A. Garcia, \textit{New regular black hole solution from nonlinear electrodynamics}, Phys. Lett. B \textbf{464} (1999) 25


\bibitem{Yajima2001}H. Yajima and Takashi Tamaki, \textit{Black hole solutions in Euler-Heisenberg theory}, Phys. Rev. D \textbf{63} (2001) 064007


\bibitem{Zumino2008}P. Aschieri, S. Ferrara, and B. Zumino, \textit{Duality rotations in nonlinear electrodynamics and in extended supergravity}, Riv. Nuovo Cim. \textbf{31} (2008) 625

\bibitem{Munoz2009}G. Muñoz and D. Tennant, \textit{Testing string theory via Born–Infeld electrodynamics?}, Phys. Lett. B, \textbf{682} (2009) 297 

\bibitem{Miskovic2011a}O. Mišković and R. Olea, \textit{Quantum statistical relation for black holes in nonlinear electrodynamics coupled to Einstein-Gauss-Bonnet AdS gravity}, Phys. Rev. D \textbf{83} (2011) 064017 

\bibitem{Miskovic2011b}O. Mišković and Rodrigo Olea, \textit{Conserved charges for black holes in Einstein-Gauss-Bonnet gravity coupled to nonlinear electrodynamics in AdS space}, Phys. Rev. D \textbf{83} (2011) 024011


\bibitem{Zhao2013}Z. Zhao, Q. Pan, S. Chen, J. Jing, \textit{Notes on holographic superconductor models with the nonlinear electrodynamics}, Nuc. Phys. B \textbf{871} (2013) 98


\bibitem{Ruffini2013}R. Ruffini,Y.-B. Wu, and S.-S. Xue, \textit{Einstein-Euler-Heisenberg theory and charged black holes}, Phys. Rev. D \textbf{88} (2013) 085004

\bibitem{Hendi2015Dehghani} S.H. Hendi and A. Dehghani, \textit{Thermodynamics of third-order Lovelock-AdS black holes in the presence of Born-Infeld type
nonlinear electrodynamics}, Phys. Rev. D \textbf{91} (2015) 064045

\bibitem{Dehyadegari2016}M. Kord Zangeneh, A. Dehyadegari, A. Sheykhi, M.H. Dehghani, \textit{Thermodynamics and gauge/gravity duality for Lifshitz black holes in the presence of exponential electrodynamics}, J. High Energy Phys. \textbf{03} (2016) 037

\bibitem{Gulin2017} L. Gulin and I. Smolić, \textit{Generalizations of the Smarr formula for black holes with nonlinear electromagnetic fields}, Class. Quant. Grav. \textbf{35} (2017) 025015




\bibitem{Falciano2019}F.T. Falciano, M. L. Peñafiel, and S.E. Perez Bergliaffa, \textit{Entropy bounds and nonlinear electrodynamics}, Phys. Rev. D \textbf{100} (2019) 125008

\bibitem{Cremonini2019} S. Cremonini, A. Hoover, L. Li, and S. Waskie, \textit{Anomalous scalings of cuprate strange metals from nonlinear electrodynamics}, Phys. Rev. D \textbf{99} (2019) 061901

\bibitem{Zarepour2021}S. Zarepour, \textit{Holographic heat engines coupled with logarithmic $U(1)$ gauge theory}, Int. J. Mod. Phys. D \textbf{30} (2021) 2150109

\bibitem{Amirabi2021}Z. Amirabi and S.H. Mazharimousavi, \textit{Black-hole solution in nonlinear electrodynamics with the maximum allowable symmetries}, Eur. Phys. J. C \textbf{81} (2021) 207 

\bibitem{Kruglov2017}S.I. Kruglov, \textit{Remarks on Heisenberg-Euler-type electrodynamics}, Mod. Phys. Lett. A \textbf{32} (2017) 1750092


\bibitem{Akmansoy2018}P.N. Akmansoy and L.G. Medeiros, \textit{Constraining Born–Infeld-like nonlinear electrodynamics using hydrogen’s ionization energy}, Eur. Phys. J. C \textbf{78} (2018) 143


\bibitem{LandauLifshitzBook} L.D. Landau and E.M. Lifshitz, \textit{The Classical Theory of Fields}, (fourth edition, Pergamon Press, Oxford U.K., 1980)


\bibitem{NED-conformalI1} J.A.R. Cembranos, A. de la Cruz-Dombriz, and J. Jarillo, \textit{Reissner-Nordström black holes in the inverse electrodynamics model}, JCAP \textbf{02} (2015) 042

\bibitem{NED-conformalI2} V.I. Denisov, E. E. Dolgaya, V.A. Sokolov, and I.P. Denisova, \textit{Conformal invariant vacuum nonlinear electrodynamics}, Phys. Rev. D \textbf{96} (2017) 036008

\bibitem{NED-conformalI3} I.P. Denisova, B.D. Garmaev, and V.A. Sokolov, \textit{Compact objects in conformal nonlinear electrodynamics}, Eur. Phys. J. C \textbf{79} (2019) 531

\bibitem{NED-conformalI4} M. Asorey, L. Rachwal, I.L. Shapiro, and W. Cesar e Silva, \textit{On the vector conformal models in an arbitrary dimension}, Eur. Phys. J. Plus \textbf{136} (2021) 1043

\bibitem{NED-conformalI5} D. Kokoska and M. Ortaggio, \textit{Static and radiating dyonic black holes coupled to conformally invariant electrodynamics in higher dimensions}, Phys. Rev. D \textbf{104} (2021) 124051

\bibitem{Ritz1996}A. Ritz and R. Delbourgo, \textit{The Low Energy Effective Lagrangian for Photon Interactions in Any Dimension}, Int. J. Mod. Phys. A \textbf{11} (1996) 253

\bibitem{Shabad2015} C.V. Costa, D.M. Gitman, and A. E. Shabad, \textit{Finite field-energy of a point charge in QED}, Phys. Scr. \textbf{90} (2015) 074012

\bibitem{Sauter1931}F. Sauter, \textit{Über das Verhalten eines Elektrons im homogenen elektrischen Feld nach der relativistischen Theorie Diracs}, Z. Physik \textbf{69} (1931) 742

\bibitem{Schwinger1951}J. Schwinger, \textit{On gauge invariance and vacuum polarization}, Phys. Rev. \textbf{82} (1951) 664


\bibitem{JacksonBook}J.D. Jackson, \textit{Classical Electrodynamics} (John Wiley \& Sons, 2007)

\bibitem{LifshitzQED}V.B. Berestetskii, E.M. Lifshitz, and L. Pitaevskii, \textit{Quantum Electrodynamics} (Pergamon Press Ltd. UK, 1980)

\bibitem{BerestetskiiQED} A.I. Akhiezer, V.B. Berestetskii, \textit{Quantum Electrodynamics}, (Interscience Publishers, New York, 1965)


\bibitem{Schrodinger1935}E. Schrödinger, \textit{Contributions to Born's new theory of the electromagnetic field}, Proc. Roy. Soc. (Lond.) A \textbf{150} (1935) 465

\bibitem{Hatsuda1999} M. Hatsuda, K. Kamimura and S. Sekiya, \textit{Electric-Magnetic Duality Invariant Lagrangians}, Nucl. Phys. B \textbf{561} (1999) 341

\bibitem{Kruglov2007}S.I. Kruglov, \textit{Vacuum birefringence from the effective Lagrangian of the electromagnetic field}, Phys. Rev. D \textbf{75} (2007) 117301

\bibitem{Rafelski1973} G. Soff, J. Rafelski, and W. Greiner, \textit{Lower bound to limiting fields in nonlinear electrodynamics}, Phys. Rev. A \textbf{7} (1973) 903





\bibitem{Enoto}
T. Enoto, K. Nakazawa, K. Makishima, N. Rea, K. Hurley, and S. Shibata, \textit{Broadband study with Suzaku of the magnetar class}, Astrophys. J. 
Lett. \textbf{722} (2010) L162

\bibitem{PDG2020}
P.A.~Zyla et al. (Particle Data Group), Prog. Theor. Exp. Phys. \textbf{2020} (2020) 083C01


%\bibitem{Shabad2011}A.E. Shabad and V.V. Usov, \textit{Effective Lagrangian in nonlinear electrodynamics and its properties of causality and unitarity}, Phys. Rev. D \textbf{83} (2011) 105006













% Please avoid comments such as "For a review'', "For some examples",
% "and references therein" or move them in the text. In general,
% please leave only references in the bibliography and move all
% accessory text in footnotes.

% Also, please have only one work for each \bibitem.


\end{thebibliography}
\end{document}